\documentclass{article}

\usepackage{PRIMEarxiv}

\usepackage[utf8]{inputenc} 
\usepackage[T1]{fontenc}    
\usepackage{hyperref}       
\usepackage{url}            
\usepackage{booktabs}       
\usepackage{amsfonts}       
\usepackage{nicefrac}       
\usepackage{microtype}      
\usepackage{lipsum}
\usepackage{fancyhdr}       
\usepackage{graphicx}       
\usepackage{float}
\usepackage{multirow}
\usepackage{subfig}
\usepackage{amsmath}
\usepackage{amssymb}
\usepackage{mathtools}
\usepackage{dcolumn}
\usepackage{bm}
\usepackage{tikz}
\usepackage{cite}
\graphicspath{{media/}}     

\newcommand{\abs}[1]{\left\lvert#1\right\rvert}

\title{Dynamics of an autocatalytic reaction front: effects of imposed turbulence and buoyancy-driven flows}

\usepackage{authblk} 
\author[1]{\textbf{Nihal Tawdi}\thanks{Email address for correspondence: nihal.tawdi@gmail.com}}
\author[1]{\textbf{Christophe Almarcha}}
\author[1]{\textbf{Michael Le Bars}}

\affil[1]{Aix Marseille Univ, CNRS, Centrale Marseille, IRPHE, Marseille, France}

\date{(Received xx; revised xx; accepted xx)}

\begin{document}

\maketitle

\begin{abstract}
	Thin flame dynamics in a turbulent flow remains debated (e.g. \cite{damkohler1940, clavin1979, klimov1983, gouldin1987, bray1990, gulder1991, sivashinsky1990}), with various parameterizations proposed for the typical flame propagation velocity. According to the classical Damköhler's model based on Huygens'  principle, a flame front should advance at a constant velocity normal to the interface between reactants and products, with a turbulent acceleration induced by the  wrinkling and surface increase of the interface. However, combustion experiments often deviate from this model \cite{karpov1980, abdelgayed1987, kido1989, bradley1992} due to intertwined thermal and turbulent effects, complicating flame acceleration characterization.
	In this study, we use an autocatalytic reaction that generates a thin reactive front in an aqueous medium, enabling clearer isolation of turbulence effects. Using oscillating grids to generate turbulence in a closed tank, we examine two configurations: a single-grid setup with a spatially decaying turbulence and a dual-grid system with in the middle, a nearly homogeneous, isotropic turbulence. Particle Image Velocimetry and Laser Induced Fluorescence measurements capture both the velocity field and the front propagation, revealing two different regimes: the expected Huygens' propagation regime, but also a reactive mixing regime, where the turbulent advection of the products inside the reactants initiates multiple, dispersed reaction locations.
	Additionally, we show that even the small density difference between reactants and products plays a crucial role in the front dynamics. This work advances our understanding of autocatalytic fronts in turbulence, emphasizing the critical interplay between chemical kinetics and flow dynamics.
\end{abstract}


\raggedbottom

\section{Introduction}\label{sec:Intro}

The interaction between a thin premixed flame and turbulence is a fundamental and complex phenomenon that requires a better understanding for various engineering applications. The non-linear correlation between the chemical kinetics and swirling eddies introduces significant complexities, making it a rich area of study for both fundamental research and practical implementation. For decades, various theoretical, numerical and experimental studies focused on finding a scaling relationship between the turbulent propagation velocity $S_{T}$, flow velocity fluctuations $u'$ describing the turbulent field, and the laminar propagation velocity $S_{L}$ characterizing the reaction-diffusion front propagation velocity  in the absence of any perturbation.
Different trends were unraveled depending on the considered hypotheses on turbulence, thermal expansion, heat release, and the intrinsic properties of the chemical reaction. As a matter of fact, in premixed combustion, thermodiffusive and hydrodynamic instabilities can occur during the flame propagation, intermingling with turbulence in establishing the front shape and velocity, making it impractical to uncouple the effect of each, and particularly to determine the specific influence of turbulence. The Huygens' model stands as the basis of empirical models of premixed turbulent flame propagation. It is based on the principle that the flame front can be visualized as a series of propagating waves, much like how Huygens' principle describes wavefronts in optics. Each point on the flame front acts as a source of spherical wavelets, which spread out to form the new flame front. Within this framework, the flame front behaves as a continuous surface, which was first described by Damköhler in 1940 \cite{damkohler1940}. He suggested that this continuous surface is wrinkled by turbulence, which increases the reaction surface and hence the mass consumption rate of the reactant, therefore speeding up the front. Each point of this wrinkled front propagates with a constant normal velocity corresponding to the laminar propagation velocity $S_{L}$. The mass balance across the interface \cite{williams1985} gives the turbulent velocity increase according to the increase in surface area
\begin{equation}
	\frac{S_{T}}{S_{L}} = \frac{A_{T}}{A_{L}},
\end{equation}
where $A_{T}$ is the surface area of the wrinkled front and $A_{L}$ is the cross-section area projected in the direction of propagation. Straightforward geometrical considerations under weak interface deformations then give a quadratic acceleration fo the front 
\begin{equation} \frac{S_{T}}{S_{L}}-1 \propto \left (\frac{u_{\text{RMS}}}{S_{L}} \right)^{2} \label{eq:CW79} \end{equation}
with $u_{\text{RMS}}$ the rms velocity. Clavin \& Williams (1979) \cite{clavin1979} derived the same dependency in a more rigorous analytical approach, using regular perturbation analysis for small $\epsilon = \frac{\delta_{L}}{\eta}$ in an advection-diffusion-reaction equation, where $\delta_{L}$ is the reaction zone thickness and $\eta$ the Kolmogorov length scale of the turbulence flow. Their analysis assumes (i) equal thermal and mass diffusivities (Lewis number $Le = 1$) to exclude thermo-diffusive instabilities and (ii) negligible density variations ($\Delta \rho / \rho$) to rule out Darrius-Landau instabilities.
These conditions ensure that flame wrinkling, driven solely by turbulence, is the dominant acceleration mechanism, assuming turbulence is weakly perturbing the flame's structure and the velocity field remains unaffected by combustion. 
Kerstein and Ashurst \cite{kerstein1992} emphasized that, in addition to deterministic stretching by the mean velocity gradients, one must also consider the nonlinear front relaxation that tends to smooth out sharp gradients, and the stochastic forcing of the front stemming from local velocity
divergence. In turbulent or random flows, this forcing acts stochastically, causing the interface to accumulate small, random distortions over time: the balance between this stochastic growth and the nonlinear relaxation then leads to a scaling of the turbulent flame speed with the rms velocity as $S_{T}/S_{L} \sim (u_{\text{RMS}}/S_{L})^{4/3}$. However, in a periodic flow, the stochastic accumulation is absent, leading back to Clavin \& Williams' \cite{clavin1979} scaling. 

In order to achieve a smooth and continuous interface between unburned and burned gases so that the front remains a connected surface (in a topological sense) and thus for the Huygens' model to stay valid, Shy \textit{et al.} (1992) \cite{shy1992} indicate that the following assumptions should hold: 
(i) a constant normal propagation velocity $S_{L}$; (ii) weak density variation $\Delta \rho$ across the interface to avoid hydrodynamic instabilities;
(iii) a statistically homogeneous and stationary turbulence; 
(iv) constant thermodynamic properties to achieve uniform transport properties across the interface. 
In the case of a combustion flame, the highly exothermic nature of the reaction gives rise to high density and temperature fluctuations leading to violation of the above assumptions. Moreover, expansion of the gases disrupts and inhomogenizes the flow, therefore complicating the study of the flame surface increase by ambient turbulence.

In order to get as close as possible to the assumptions stated above, and to avoid any hydrodynamic, thermodiffusive or thermo-acoustic instabilities, Ronney \cite{ronney1995} introduced the concept of \textit{liquid flames}. A liquid flame is given by an autocatalytic chemical reaction in an aqueous (hence incompressible) medium, which gives rise -- analogously to a premixed flame in combustion -- to a thin front at the interface between reactants and products, where the reaction takes place. In an autocatalytic chemical reaction, one of the products also serves as a catalyst: the reaction is self-sustained until the limiting reagent is exhausted, just as the heat released by a combustion reaction allows the flame to propagate in the fuel and oxidizer mixture. Indeed, while the heat released by the combustion process locally rises the temperature at the flame location, which causes the fresh gases to react, the catalyst production at the front in the autocatalytic reaction locally initiates the reaction, thus propagating the front. \\
The present study follows the previous works by Ronney, Shy, and collaborators \cite{shy1992,shy1996a,shy1996b,shy1997,shy1999}, using the same arsenous-acid reaction as well as an oscillating grid apparatus to generate turbulence. While these previous studies focused solely on the liquid flame propagation in the homogeneous and isotropic turbulence created by a two oscillating grids system, we show how this particular setup does not exclusively give rise to front propagation following the Huygens' approximation. This is revealed by using a single oscillating grid, located either in the products or in the reactants. Our investigation elucidates the consequence of the asymmetry between products and reactants from a chemodynamics point of view, highlighting the  existence of two separate regimes: 
\begin{itemize}
	\item the expected propagation regime when the oscillating grid is located in the products,  where products and reactants remain separated by a sharp interface evolving from products towards reactants. The front shape, and hence propagation velocity, depend on the local turbulence intensity, which decreases when going farther away from the grid, resulting in a time and space decreasing front velocity.
	\item a new reactive mixing regime when the oscillating grid is located in the reactants, where the reaction does not proceed with a continuous front but rather at multiple, dispersed locations within the medium. 
\end{itemize}

\begin{figure}
	\centering
	\includegraphics[width=0.7\linewidth]{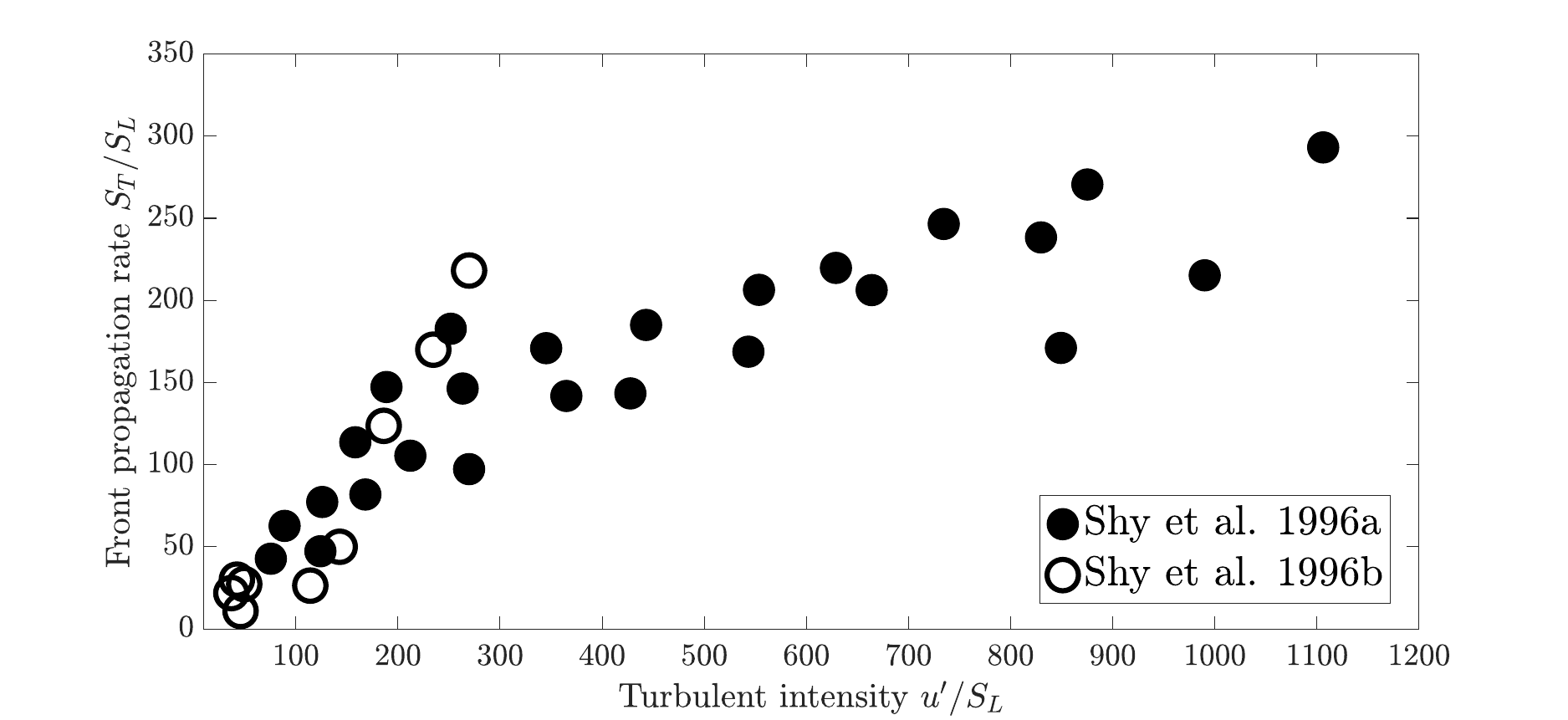}
	\caption{Evolution of the turbulent front propagation rate $S_{T}/S_{L}$ as a function of the turbulent intensity rate $u'/S_{L}$ found by Shy \textit{et al.} \cite{shy1996a,shy1996b}.}
	\label{fig:shy1996}
\end{figure}

The coexistence and competition between both regimes when oscillating 2 grids simultaneously might contribute to explain the rather large results dispersion observed in previous studies and illustrated in figure \ref{fig:shy1996}, with no general agreement with the Huygens' propagation model. 
Furthermore, by exploring an extended range of reactants concentration, our investigation exhibits the importance of buoyancy effects, up to now neglected. While the density difference ${\Delta \rho}/{\rho}$ between products and reactants, typically of order $0.01\%$, was initially considered  to have a negligible effect on the front propagation, one grid experiments unravel a clear feedback of the front propagation on the surrounding flow due to the buoyancy forces applied to the stretched interface. From this, a relationship between $S_{T}/S_{L}$, $u_{\text{RMS}}/S_{L}$, and the Froude number $Fr$, is derived and discussed.

Our paper is organized as follows. Section 2 presents the experimental apparatus, including the considered chemical reaction and the metrology. Section 3 examines the case of a single oscillating grid, exploring spatially decreasing turbulence and characterizing the two regimes introduced above. Section 4 then revisits the case of two oscillating grids, where homogeneous and isotropic turbulence is achieved, and highlights the previously overlooked influence of chemical density stratification. Finally, Section 5 concludes the paper and discusses prospects for future research.

\section{Experimental apparatus}

\subsection{Iodate -- Arsenous Acid reaction}

To obtain a sharp diffusion-reaction front, the Iodate -- Arsenous Acid (IAA) reaction is investigated, as previously done without \cite{hanna1982,pojman1991,baba2018,saul1985} or with an imposed background flow \cite{shy1992,shy1996a,shy1996b,shy1999}. 
The reaction between arsenous acid $H_{3}AsO_{3}$ and iodate ions $IO_{3}^{-}$, produces arsenic acid $H_{3}AsO_{4}$ and iodide ions $I^{-}$ provided that the ratio between the initial reactants concentrations $R=\frac{[H_{3}AsO_{3}]_{0}}{[IO_{3}^{-}]_{0}}$ is greater than or equal to 3, \textit{i.e.} that arsenous acid is in stoichiometric excess. The reaction writes  
\begin{equation}
	3H_{3}AsO_{3} + IO_{3}^{-} + 5I^{-}\rightarrow 3H_{3}AsO_{4} + 6I^{-}.
	\label{eq:reaction_globale}
\end{equation}
The IAA reaction is an autocatalytic reaction, meaning it produces its own catalyst, which is iodide $I^{-}$ in this case. This reaction can be divided into two intermediate steps: 
\begin{itemize}
	\item 
	the Dushman reaction describing the oxidation of iodide by iodate 
	\begin{equation}
		IO_{3}^{-} + 5I^{-} + 6H^{+} \rightarrow 3I_{2} + 3H_{2}O,
		\label{eq:dushman}  
	\end{equation}
	\item 
	followed by the Roebuck reaction reducing iodine by arsenous acid  
	\begin{equation}
		H_{3}AsO_{3} + I_{2} + H_{2}O \rightarrow 2I^{-} + H_{3}AsO_{4} + 2H^{+},
		\label{eq:roebuck}    
	\end{equation}
\end{itemize}
such that \eqref{eq:dushman}$+3 \times$\eqref{eq:roebuck}=\eqref{eq:reaction_globale}.
Reaction \eqref{eq:dushman} consumes iodide $I^{-}$ and produces iodine $I_{2}$, which is then reduced by the more rapid reaction $\eqref{eq:roebuck}$ to regenerate iodide, the catalyst. As a result, the iodide concentration $[I^{-}]$ increases autocatalytically until the iodate $IO_{3}^{-}$, i.e. the limiting reactant, is fully consumed.

\begin{figure}[H]
	\centering
	\begin{minipage}{0.2\textwidth}
		\centering
		\includegraphics[width=0.5\textwidth]{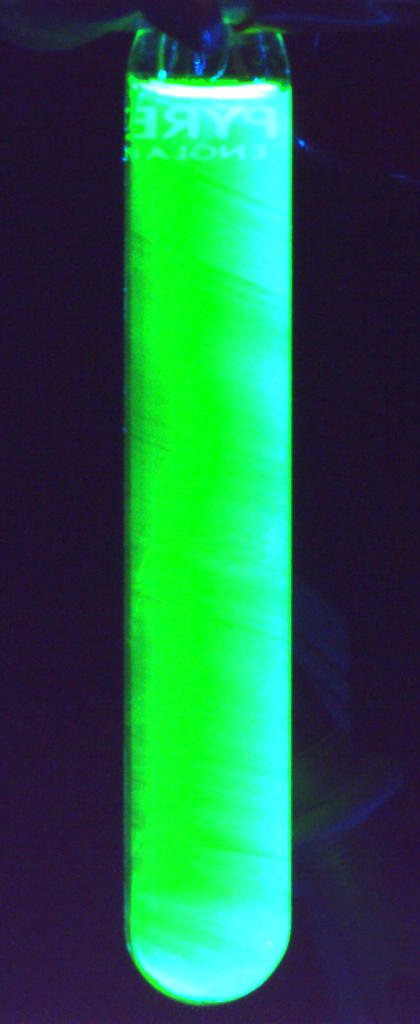}\\
		\scriptsize (a) $t=0$
	\end{minipage}%
	\hspace{-0.8cm}
	\begin{minipage}{0.2\textwidth}
		\centering
		\includegraphics[width=0.5\textwidth]{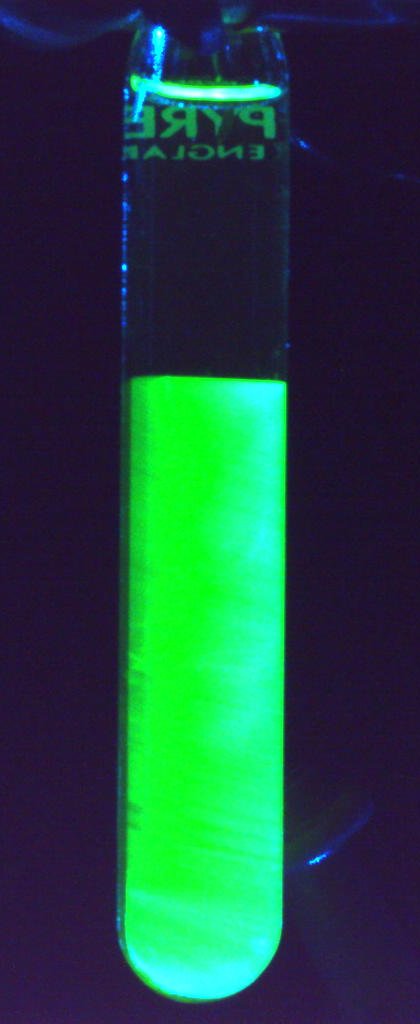}\\
		\scriptsize (b) $t=29$~min~30~s.
	\end{minipage}%
	\hspace{-0.8cm}
	\begin{minipage}{0.2\textwidth}
		\centering
		\includegraphics[width=0.5\textwidth]{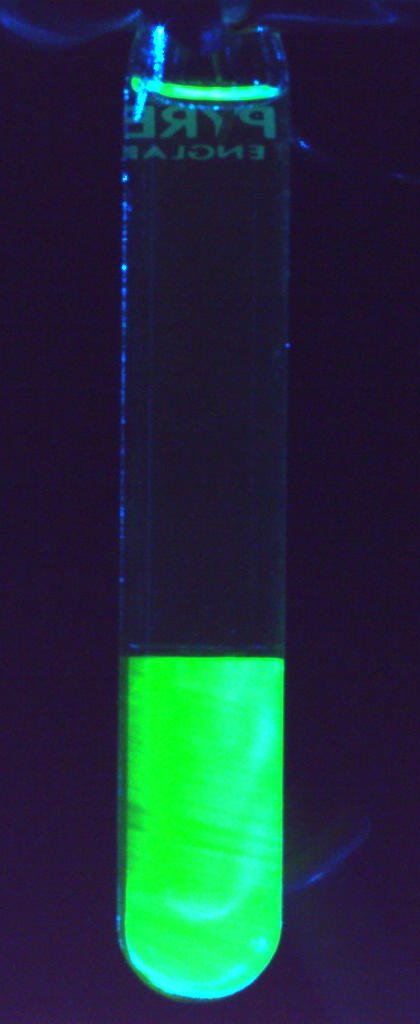}\\
		\scriptsize (c) $t=58$~min~28~s.
	\end{minipage}%
	\hspace{-0.8cm}
	\begin{minipage}{0.2\textwidth}
		\centering
		\includegraphics[width=0.5\textwidth]{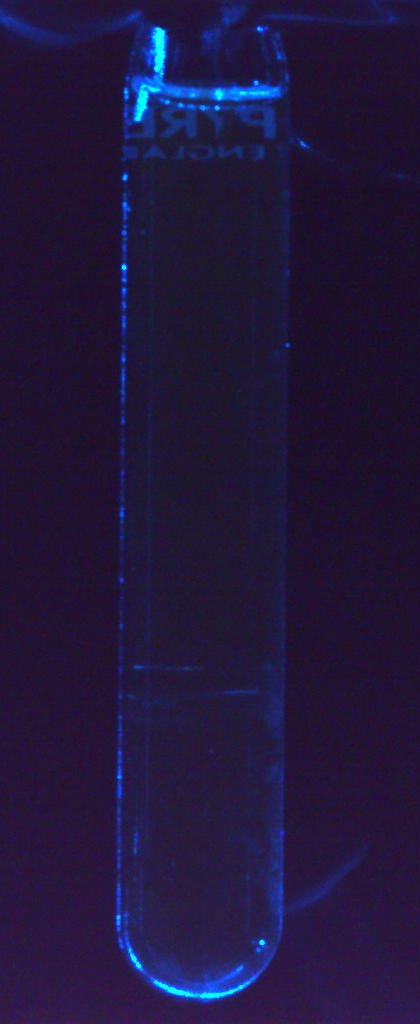}\\
		\scriptsize (d) $t=1$~h~32~min~44~s.
	\end{minipage}
	\caption{Snapshots of the front propagation in a 15$\times$100 mm tube seen from the side at different times, starting with reaction initiation at $t=0$. In this experiment, $[H_{3}AsO_{3}]_{0} = 0.05$ mol/l, $[IO_{3}^{-}]_{0}=0.017$ mol/l ($R=3$) and the initial pH was set to $pH_{i}$  = 8.6. The front is visualized with fluorescein, a pH-sensitive dye that fluoresces in the reactants where $pH > 4$.}\label{fig:tube_time_evolution}
\end{figure}

This reaction, in the absence of stirring, gives rise to a sharp front made of a thin reaction zone where intermediate reactions take place, separating the reactants that are disappearing as the front is progressing from the products that replace them. Figure \ref{fig:tube_time_evolution} shows an example of a propagating front, where reactants appear in green and products in black. This visual separation, which allows the front to be tracked, is obtained by laser induced fluorescence (LIF) thanks to fluorescein $C_{20}H_{12}O_{5}$,  which is a highly fluorescent and water-soluble dye that emits green when under blue light, but only when the pH of the dyed solution is above 4. An interesting feature of the IAA reaction is that while the reactants are neutral or slightly basic with a pH between 7 and 8, the products are more acidic, with a pH around 2. 

A slight density difference $\Delta \rho$ of order $10^{-4}$ g.cm$^{-3}$ between products and reactants has been previously reported  \cite{rongy2007buoyancy,pojman1991,popity2013}, such that $\rho_{\text{products}} < \rho_{\text{reactants}}$. Therefore, downward front propagation is hydrodynamically stable, with an horizontal interface separating the less dense products  above the denser reactants. Then, in the absence of any external perturbation, the propagation is only driven by diffusion and reaction, and the front velocity is referred to as the laminar propagation velocity. 
The influence of different parameters such as reactants initial concentrations $[H_{3}AsO_{3}]_{0}$ and $[IO_{3}^{-}]$ as well as the initial $pH$ and temperature $T$ on the reaction rate, hence on the laminar propagation velocity, has been extensively studied in the literature \cite{hanna1982,valkai2016,mercer2007}.

\subsection{Turbulence generation apparatus}

\vspace{-1cm}
\begin{figure}[H]
	\centering
	\subfloat[Photo]{\includegraphics[width=0.25\textwidth]{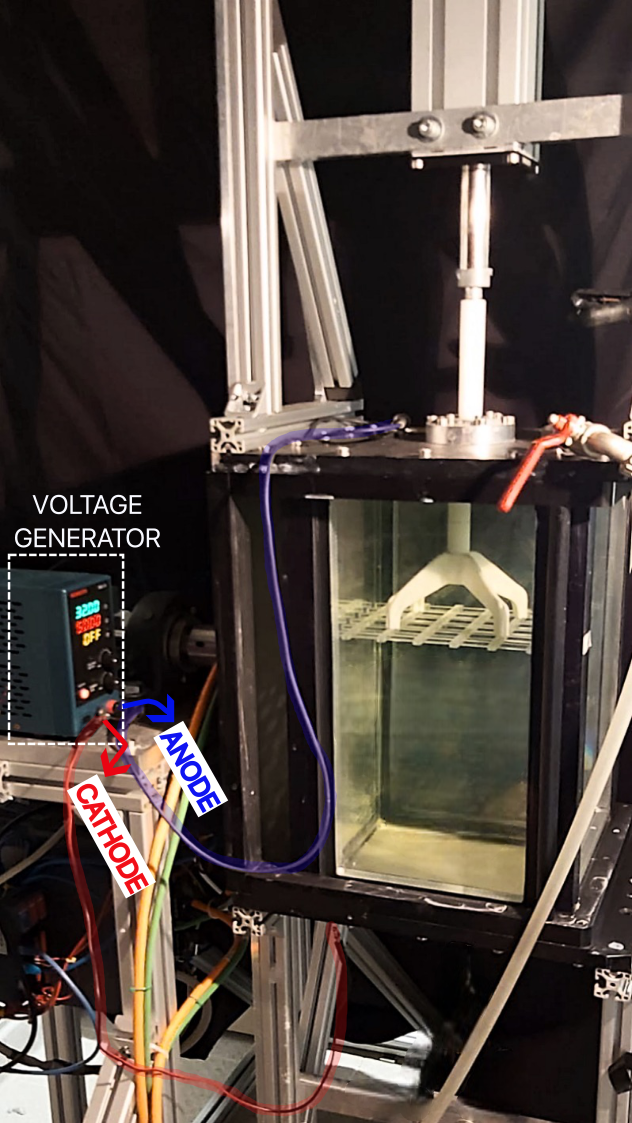}\label{fig:init}} 
	\subfloat[Sketch]{\includegraphics[width=0.7\textwidth]{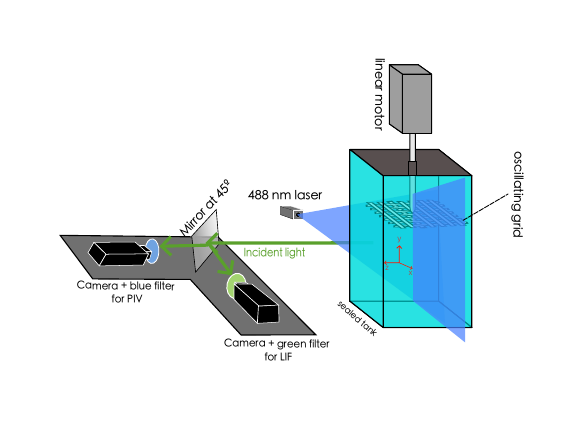}}
	\caption{Experimental apparatus in the one-grid configuration. In the photo (a), the anode, connecting the positive pole of the generator to the top of the tank, is highlighted in blue and the cathode, connecting the negative pole of the generator to the bottom of the tank, is highlighted in red. In the sketch (b), the dual mount system for PIV and LIF synchronized cameras is illustrated.}
	\label{fig:PhotoManip}
\end{figure}

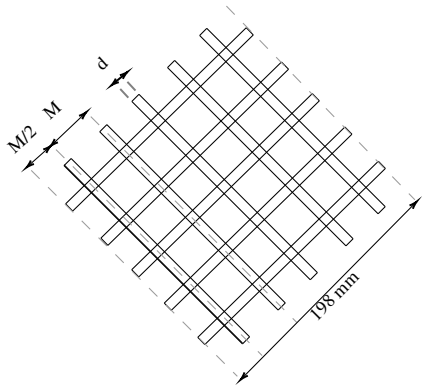
\begin{figure}[H]
	\centering
	\begin{tikzpicture}[x=0.75pt,y=0.75pt,yscale=-0.9,xscale=0.9]
		
		\draw    (100,121) -- (200,221) ;
		\draw    (104,117) -- (204,217) ;
		\draw    (100,121) -- (104,117) ;
		\draw    (200,221) -- (204,217) ;
		\draw    (120,102) -- (220,202) ;
		\draw    (124,98) -- (224,198) ;
		\draw    (120,102) -- (124,98) ;
		\draw    (220,202) -- (224,198) ;
		\draw    (138,85) -- (238,185) ;
		\draw    (142,81) -- (242,181) ;
		\draw    (138,85) -- (142,81) ;
		\draw    (238,185) -- (242,181) ;
		\draw    (158,66) -- (258,166) ;
		\draw    (162,62) -- (262,162) ;
		\draw    (158,66) -- (162,62) ;
		\draw    (258,166) -- (262,162) ;
		\draw    (176,48) -- (276,148) ;
		\draw    (180,44) -- (280,144) ;
		\draw    (176,48) -- (180,44) ;
		\draw    (276,148) -- (280,144) ;
		\draw    (202.11,43.65) -- (100.9,142.42) ;
		\draw    (206.06,47.7) -- (104.85,146.47) ;
		\draw    (202.11,43.65) -- (206.06,47.7) ;
		\draw    (100.9,142.42) -- (104.85,146.47) ;
		\draw    (221.87,62.88) -- (120.65,161.65) ;
		\draw    (225.82,66.93) -- (124.6,165.7) ;
		\draw    (221.87,62.88) -- (225.82,66.93) ;
		\draw    (120.65,161.65) -- (124.6,165.7) ;
		\draw    (239.14,80.59) -- (137.93,179.36) ;
		\draw    (243.09,84.64) -- (141.88,183.41) ;
		\draw    (239.14,80.59) -- (243.09,84.64) ;
		\draw    (137.93,179.36) -- (141.88,183.41) ;
		\draw    (257.4,99.32) -- (156.18,198.09) ;
		\draw    (261.35,103.37) -- (160.13,202.14) ;
		\draw    (257.4,99.32) -- (261.35,103.37) ;
		\draw    (156.18,198.09) -- (160.13,202.14) ;
		\draw    (276.15,118.53) -- (174.94,217.3) ;
		\draw    (280.1,122.58) -- (178.89,221.35) ;
		\draw    (276.15,118.53) -- (280.1,122.58) ;
		\draw    (174.94,217.3) -- (178.89,221.35) ;
		\draw [color={rgb, 255:red, 155; green, 155; blue, 155 }  ,draw opacity=1 ] [dash pattern={on 4.5pt off 4.5pt}]  (95.5,112) -- (211,227.5) ;
		\draw [color={rgb, 255:red, 155; green, 155; blue, 155 }  ,draw opacity=1 ] [dash pattern={on 4.5pt off 4.5pt}]  (115,93.5) -- (230.5,209) ;
		\draw    (111.05,92.38) -- (93.95,108.62) ;
		\draw [shift={(92.5,110)}, rotate = 316.47] [color={rgb, 255:red, 0; green, 0; blue, 0 }  ][line width=0.75]    (4.37,-1.32) .. controls (2.78,-0.56) and (1.32,-0.12) .. (0,0) .. controls (1.32,0.12) and (2.78,0.56) .. (4.37,1.32)   ;
		\draw [shift={(112.5,91)}, rotate = 136.47] [color={rgb, 255:red, 0; green, 0; blue, 0 }  ][line width=0.75]    (4.37,-1.32) .. controls (2.78,-0.56) and (1.32,-0.12) .. (0,0) .. controls (1.32,0.12) and (2.78,0.56) .. (4.37,1.32)   ;
		\draw    (100,120.5) -- (200,220.5) ;
		\draw [color={rgb, 255:red, 128; green, 128; blue, 128 }  ,draw opacity=1 ][line width=0.75]  [dash pattern={on 4.5pt off 4.5pt}]  (131.5,78) -- (138,85) ;
		\draw [color={rgb, 255:red, 128; green, 128; blue, 128 }  ,draw opacity=1 ][line width=0.75]  [dash pattern={on 4.5pt off 4.5pt}]  (135.5,74) -- (142,81) ;
		\draw [line width=0.75]    (133.96,69.78) -- (128.04,74.72) ;
		\draw [shift={(126.5,76)}, rotate = 320.19] [color={rgb, 255:red, 0; green, 0; blue, 0 }  ][line width=0.75]    (4.37,-1.32) .. controls (2.78,-0.56) and (1.32,-0.12) .. (0,0) .. controls (1.32,0.12) and (2.78,0.56) .. (4.37,1.32)   ;
		\draw [shift={(135.5,68.5)}, rotate = 140.19] [color={rgb, 255:red, 0; green, 0; blue, 0 }  ][line width=0.75]    (4.37,-1.32) .. controls (2.78,-0.56) and (1.32,-0.12) .. (0,0) .. controls (1.32,0.12) and (2.78,0.56) .. (4.37,1.32)   ;
		\draw [color={rgb, 255:red, 155; green, 155; blue, 155 }  ,draw opacity=1 ] [dash pattern={on 4.5pt off 4.5pt}]  (81.63,123.66) -- (197.13,239.16) ;
		\draw    (91.01,111.33) -- (80.49,120.67) ;
		\draw [shift={(79,122)}, rotate = 318.37] [color={rgb, 255:red, 0; green, 0; blue, 0 }  ][line width=0.75]    (4.37,-1.32) .. controls (2.78,-0.56) and (1.32,-0.12) .. (0,0) .. controls (1.32,0.12) and (2.78,0.56) .. (4.37,1.32)   ;
		\draw [shift={(92.5,110)}, rotate = 138.37] [color={rgb, 255:red, 0; green, 0; blue, 0 }  ][line width=0.75]    (4.37,-1.32) .. controls (2.78,-0.56) and (1.32,-0.12) .. (0,0) .. controls (1.32,0.12) and (2.78,0.56) .. (4.37,1.32)   ;
		\draw [color={rgb, 255:red, 155; green, 155; blue, 155 }  ,draw opacity=1 ] [dash pattern={on 4.5pt off 4.5pt}]  (191,31.5) -- (296.63,140.16) ;
		\draw    (296.22,142.58) -- (200.41,239.58) ;
		\draw [shift={(199,241)}, rotate = 314.65] [color={rgb, 255:red, 0; green, 0; blue, 0 }  ][line width=0.75]    (4.37,-1.32) .. controls (2.78,-0.56) and (1.32,-0.12) .. (0,0) .. controls (1.32,0.12) and (2.78,0.56) .. (4.37,1.32)   ;
		\draw [shift={(297.63,141.16)}, rotate = 134.65] [color={rgb, 255:red, 0; green, 0; blue, 0 }  ][line width=0.75]    (4.37,-1.32) .. controls (2.78,-0.56) and (1.32,-0.12) .. (0,0) .. controls (1.32,0.12) and (2.78,0.56) .. (4.37,1.32)   ;
		
		\draw (85.72,90.37) node [anchor=north west][inner sep=0.75pt]  [rotate=-317.97] [align=left] {{\scriptsize M}};
		\draw (114.72,62.87) node [anchor=north west][inner sep=0.75pt]  [rotate=-317.97] [align=left] {{\scriptsize d}};
		\draw (65.72,108.87) node [anchor=north west][inner sep=0.75pt]  [rotate=-317.97] [align=left] {{\scriptsize M/2}};
		\draw (234.63,203.45) node [anchor=north west][inner sep=0.75pt]  [rotate=-314.92] [align=left] {{\scriptsize 198 mm}};

	\end{tikzpicture}
	\caption{Grid dimensions.}
	\label{fig:dessin_grilles}
\end{figure}

Our experimental system is shown in figure \ref{fig:PhotoManip}. The closed sealed tank measures 200 $\times$ 200 $\times$ 400 mm  and three sides of the tank are made of glass for viewing the interior, with double-glazing for thermal insulation. The oscillating grid (see also figure \ref{fig:dessin_grilles}) has a mesh size $M$ = 33 mm and a bar thickness $d$ = 6.67 mm with a square cross-section, and is mounted on the shaft of a linear motor. The latter is controlled via an in-house built LabView interface, which takes as input parameters the amplitude $A$ and the frequency $f$ to form the sinusoidal signal for the position of the grid $x(t) = -A\sin(2\pi ft)$. 
A 1W blue (488 nm) laser illuminates the center plane of the tank perpendicular to the grid thanks to a Powell lense that transforms the laser beam into a 1 mm thick laser sheet.
For flow and front visualization, two synchronized cameras capture the same field of view with a resolution up to $2560 \times 2160$ pixels and  a maximum frame rate of 50 fps. For LIF measurements, a green filter is placed in front of the first camera for front tracking.  For PIV measurements, a blue filter is placed in front of the second camera to visualize the seeding particles that reflect the blue light from the laser. We use silver-coated hollow glass spheres with an average diameter of 10 $\mu m$.
PIV computations were performed with DPIV Soft \cite{DPIVSoft} using PIV boxes of size $32 \times 32$ px$^{2}$ and an overlap of 50\%. \\

\label{section:geometricalrequirements}
In order to avoid producing mean-flow and/or a shear-driven flow in the tank, previous studies have identified several conditions to be met with regard to the geometry of the grid and its positioning, namely:
\begin{itemize}

	\item to avoid edge effects, the grid must be positioned at a distance $z_{0}$ from the bottom or top of the tank such that $\frac{z_{0}}{M} > $2.5 \cite{hopfinger1975}. To achieve this, a distance of at least 100 mm between the grids and the walls is satisfied for each experiment.
	\item the solidity $\sigma = \frac{d}{M}\left(2-\frac{d}{M}\right)$ of the grid must be less than 40\% so that the individual jets produced beyond the bars do not deviate from their central axis at the risk of merging to form wider jets producing shear which reinforces the anisotropic shear-induced turbulent flux \cite{corrsin1963}. The grid therefore respects a ratio of $\frac{M}{d} = 5$, giving a solidity $\sigma$ = 36\%.
	\item to avoid large-scale secondary flows induced by the gap between the vessel and the bars at the side of the grid, the sidewall -- first bar distance must be  equal to $\frac{M}{2}$ \cite{fernando1993}. Actually, to avoid grid rubbing against the walls, an additional spacing of 1 mm was allowed between each edge of the grid and the walls.
	
	\item for the dissipation spectrum zone to be sufficiently far from the production spectrum zone, the Reynolds number must be greater than $500$, even  only 300 according to \cite{bouvard1967}. Taking $M$ = 33 mm, $u = fA$ the speed of the grid oscillations, with $f \in [3; 7]$ Hz and $A \in [5; 40]$ mm, $\nu$ the viscosity of water, we find $Re \in [495; 9240]$.
\end{itemize}
Cases with only a top grid (as illustrated in figure \ref{fig:PhotoManip}), with only a bottom grid, and with both the top and bottom grids oscillating in phase opposition, were all investigated. For all experiments presented in the following, the amplitude $A=10$ mm was set to minimize large scale recirculation in the tank. To vary turbulence intensity, only the frequency was changed in the range $[2;7]$ Hz. 

\subsection{Experimental protocol}

\begin{table}[H]
	\begin{center}
		\renewcommand{\arraystretch}{1.5}
		\begin{tabular}{|c|c|c|c|c|c|c|c|c|}
			\hline
			\multirow{2}{*}{Solution} & Volume& $[H_{3}AsO_{3}]_{0}$ &$[IO_{3}^{-}]_{0}$ & $[SO_{4}^{2-}]_{0}$& $[H_{2}PO_{4}^{-}] $&$[HPO_{4}^{2-}] $ & $S_{L}$ & $\Delta \rho/\rho$\\ 
			& (l) & (mol/l) & (mol/l) & (mol/l) & (mol/l) & (mol/l) & (mm/min) & (\%) \\
			\hline
			
			\textbf{SOL1} & \multirow{2}{*}{16}& $5\cdot10^{-2}$ & $7\cdot10^{-3}$ & $2.5\cdot10^{-2}$ & -- &--  & $1.6 \pm 0.03$ & 0.07 \\ \cline{1-1}\cline{3-9}
			
			\textbf{SOL2} &  & $10^{-2}$ & $3.3\cdot 10^{-3}$ & $5 \cdot 10^{-3}$ & $8\cdot10^{-4}$  & $2\cdot10^{-3}$ & $0.3 \pm 0.02$ & 0.05 \\ \hline
			
		\end{tabular}
		\vspace{0.5cm}
		\caption{Chemical parameters of both studied reactive solutions in this study. Here, the dihydrogen phosphate/monohydrogen phosphate ($H_{2}PO_{4}^{-}/HPO_{4}^{2-}$) pair is employed in SOL2 to precisely control the initial pH of the reactants, enabling modulation of the laminar propagation velocity across an order of magnitude, while remaining chemically inert with respect to the reactants.}
		
		\label{tab:finalSolutions}
	\end{center}
\end{table}
\vspace{-1cm}
In our study, the primary criterion for selecting initial concentrations was the transparency of the product solution in order to be able to clearly see PIV particles. In fact, the higher  the initial concentrations, the darker was the solution and the less shiny were the PIV particles inside. Two different reactive solutions were investigated, as presented in table \ref{tab:finalSolutions}. The use of the phosphate pair in \textbf{SOL2} lowers the initial pH of the reactants, thereby further reducing the laminar propagation velocity in addition to the decrease resulting from the reduced reactant concentrations, yielding a fivefold reduction in $S_L$ compared with \textbf{SOL1}.
Each reactant was prepared separately the day before performing an experiment and stirred the whole night in the dark in 8.5 liters of distilled water each. They were mixed the next day, then fluorescein and PIV particles were added to the mixture. The 16 liters tank was filled by gravity, placing the bucket containing the mixture a few centimeters above it. Once the tank was full -- with the precautionary additional liter still in the bucket and the tube linking it to the tank -- the mixture was stirred vigorously inside the tank using the grid system for 10 minutes for better homogeneity of the reactive mixture. Then, each grid was placed at its initial position and kept still. Two platinum wires were dipped in the mixture through sealed holes, one at the bottom and the other at the top of the tank. As illustrated in figure \ref{fig:init}, the upper wire, the anode, was connected to the positive pole (+) of a voltage generator while the lower one, the cathode, was connected to the negative pole (-). To start the reaction, 10 Volts were switched on for about 30 seconds. It is important that the anode, and therefore the initiation, was placed at the top of the tank, otherwise less dense products would buoyantly rise, hence creating a large convective motion inside the tank and preventing the stable, planar front from forming. On the other hand, when the reaction was initiated at the top, even from a single point, a sharp and flat interface separating products from reactants formed after a short transient time. 
For experiments with the top grid involved, we then waited for the laminar propagation velocity to bring the planar front at least one $M$ below the top grid to avoid the high shear next to the grid. 	
At the end of each experiment, the tank was completely filled with catalyst-rich product. Therefore, it had to be thoroughly cleaned before the next experiment, as even a single drop of product left behind could initiate a reaction when fresh reactants were added.

\section{Front propagation in spatially decaying turbulence}

\subsection{Velocity profile generated by a single oscillating grid}

Thompson \& Turner (1975) \cite{thompson1975} suggested that for a homogeneous (in the planes parallel to the grid) and statistically steady turbulence continuously generated by a grid oscillating around $y=0$, the typical turbulent velocity decreases with the distance to the grid $y$ as 
\begin{equation}
	u = u_{*}\left ( \dfrac{y}{y_{*}}\right )^{\frac{-B}{3\beta}}, 
	\label{eq:loi_decroissance_th}
\end{equation}
where $\beta$ denotes the proportionality coefficient between the integral length scale $l$ and the distance to the grid $y$ \cite{hopfinger1975} and $B$ is derived from the prefactor that relates the dissipation rate $\epsilon$ to $u^{3}/l$. $y_*$ is introduced here for adimensionalisation, and $u_*$ is the velocity at $y=y_*$. Note that 
this solution diverges when $y \rightarrow 0$, so introducing a ``virtual origin'' $y_{0}<0$ is needed, so that the kinetic energy actually remains finite at the grid location. 

\begin{figure}[H]
	\centering
	\includegraphics[width=0.8\textwidth]{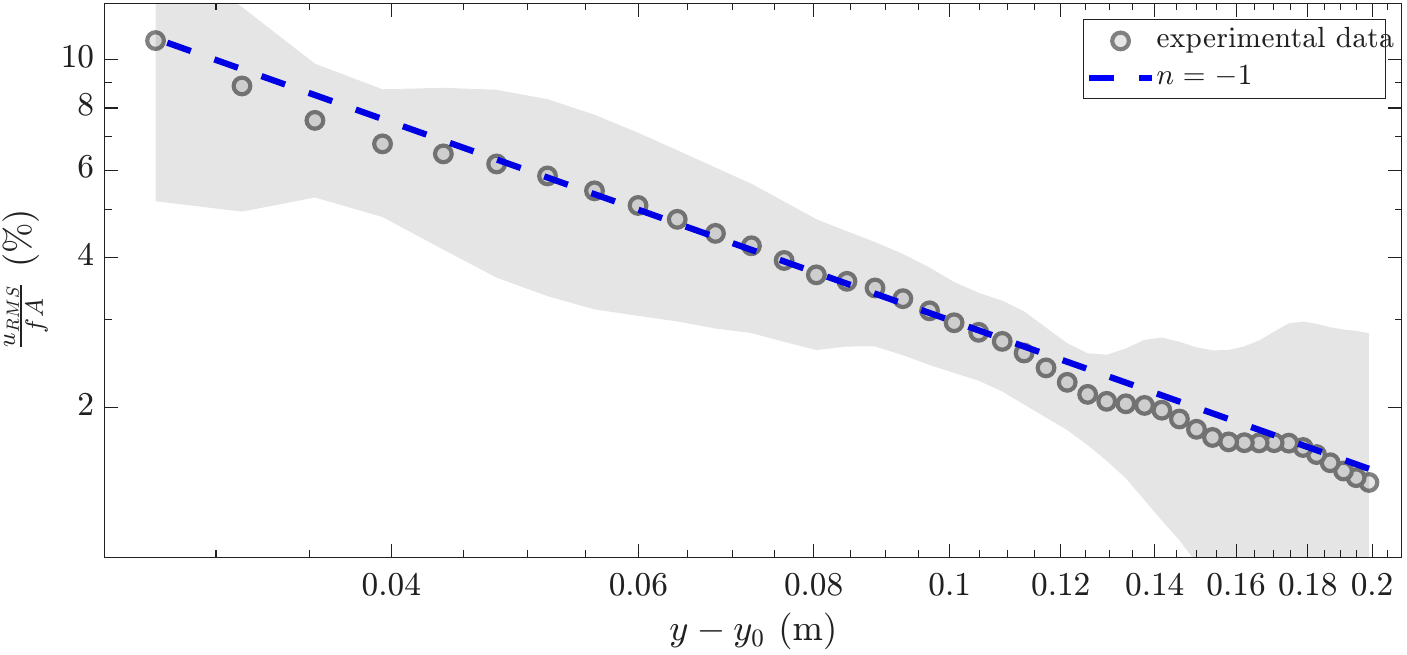}
	\caption{Evolution of $u_{RMS}(y)$ non-dimensionalized by the grid velocity $fA$ in \% along the vertical direction. The position $y=0$ is here located at the lowest position of the grid. The dashed line corresponds to $CA^{1/2}M^{1/2}(y-y_{0})^{-1}$ and the circles are the experimental measurements for $f=6$ Hz and $A=10$ mm. The shaded area shows ±1 standard deviation across the $\Vec{x}$-direction.}
	\label{fig:decayLaw_1g}
\end{figure}

Hopfinger \& Toly (1976) \cite{hopfinger1975} suggested that for square mesh grids, both $Re_{\lambda} = \frac{u_{\text{RMS}}\lambda}{\nu}$ and the eddy coefficient $K_{E}=\frac{l}{\lambda}$, where $\lambda$ is the Taylor microscale, are constant. This leads to $Re = \frac{u_{\text{RMS}}l}{\nu} = Re_{\lambda}K_{E}$ constant. Hence, accounting for (\ref{eq:loi_decroissance_th}) and for the proportionality between $y$ and $l$, $B/3\beta = 1$.
Then, the following power law, depending on $f$ the frequency of oscillations (i.e. the unique timescale in the inviscid system), $A$ their amplitude, and $M$ the mesh size, was experimentally found by \cite{hopfinger1975} and retrieved in our experimental setup 
\begin{equation}
	u_{RMS}(y) = CA^{3/2}M^{1/2}f(y-y_0)^{-1},
	\label{eq:loi_hopfinger}
\end{equation}
with $C$ a constant. In figure \ref{fig:decayLaw_1g},  the spatial decay of $u_{RMS}(y)$ agrees pretty well with \eqref{eq:loi_hopfinger}.
Here, $C=1.65\times10^{1}$ and the virtual origin was found at $y_{0}=-0.02$ m for the best fit. 

\subsection{Propagation regime}
\label{section:propagationregime}

We now focus on the case where only the top grid oscillates in the less dense products. The first regime is then highlighted, referred to as the propagation regime. 
\vspace{-0,5cm}
\begin{figure}[H]
	\centering
	\subfloat[$t= 35$s.]{\includegraphics[width=0.35\textwidth]{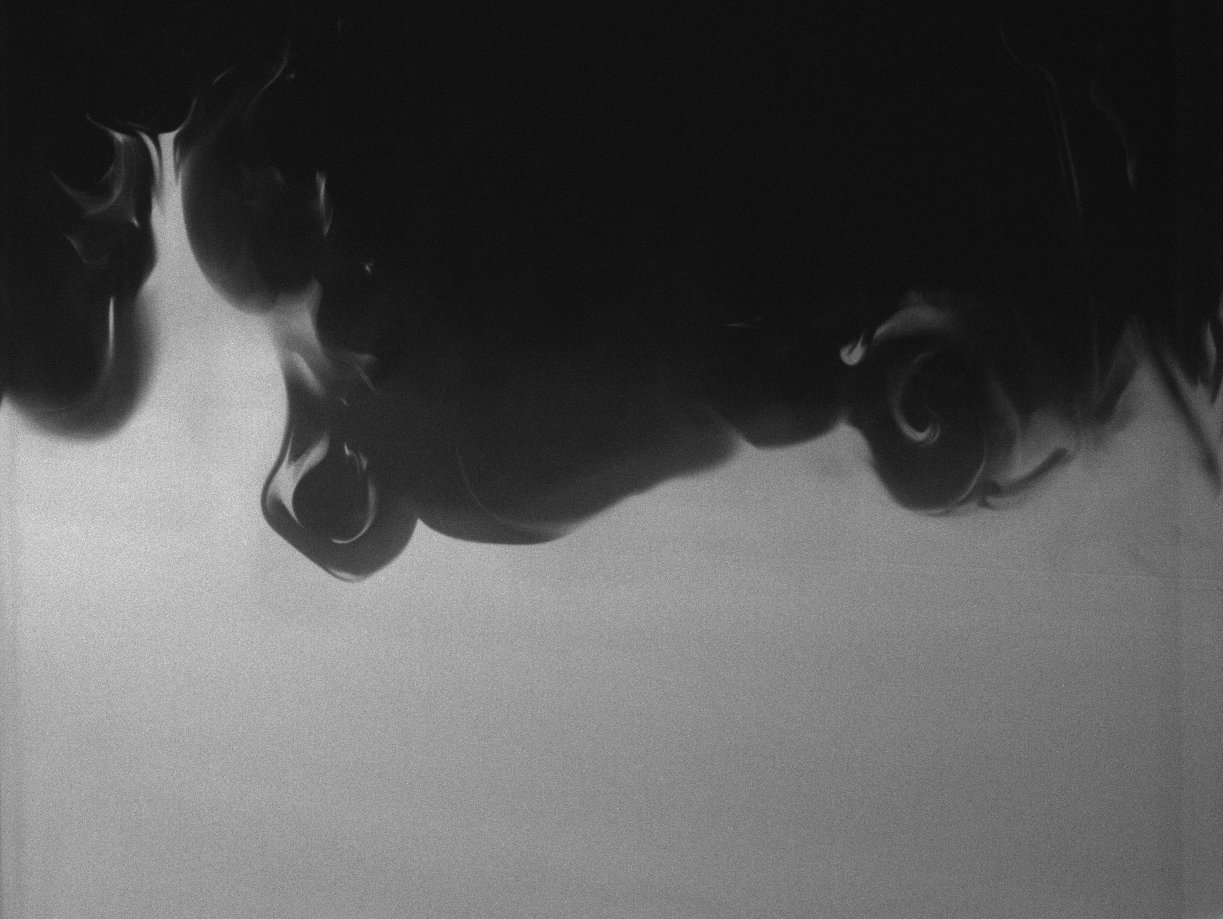}} \quad
	\subfloat[$t= 64$ s.]{\includegraphics[width=0.35\textwidth]{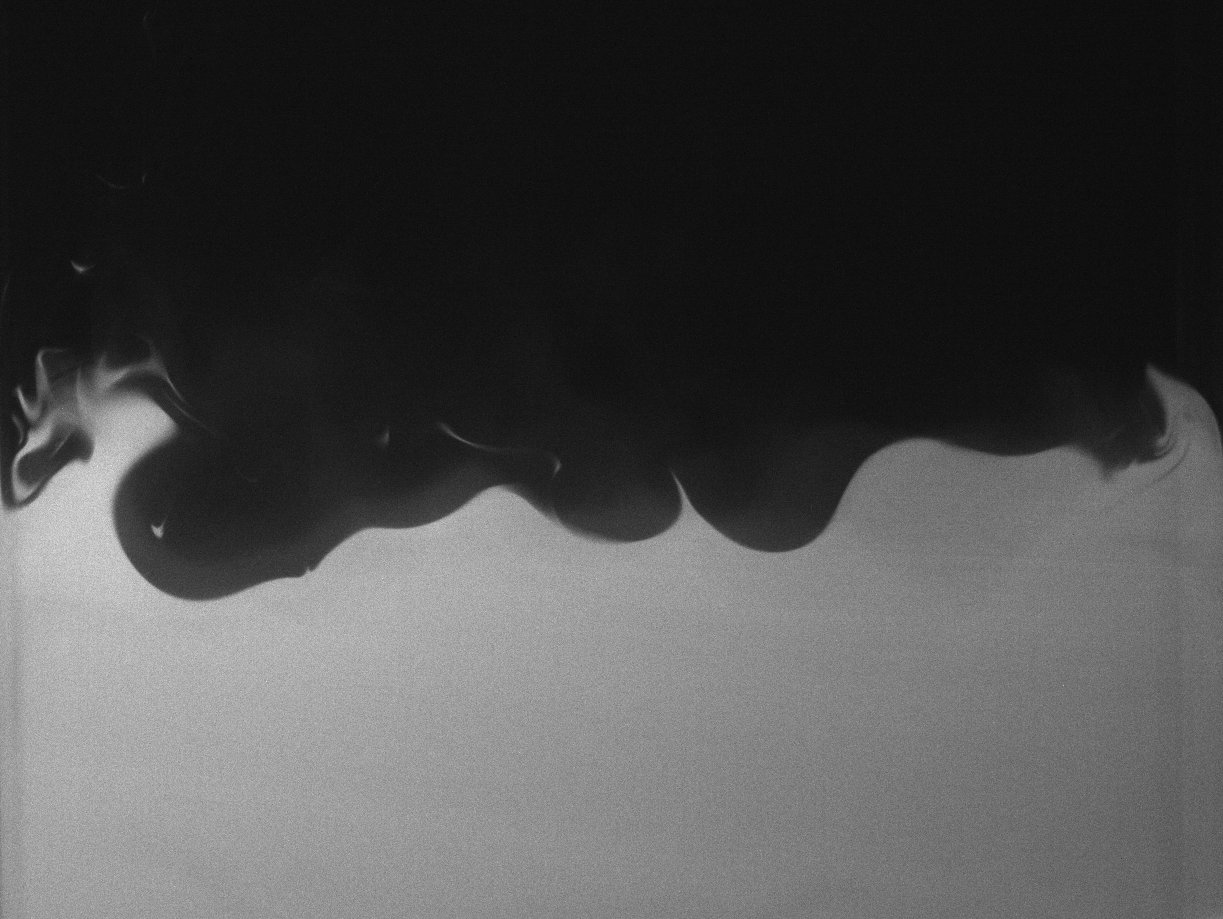}} \\
	\subfloat[$t=302$ s.]{\includegraphics[width=0.35\textwidth]{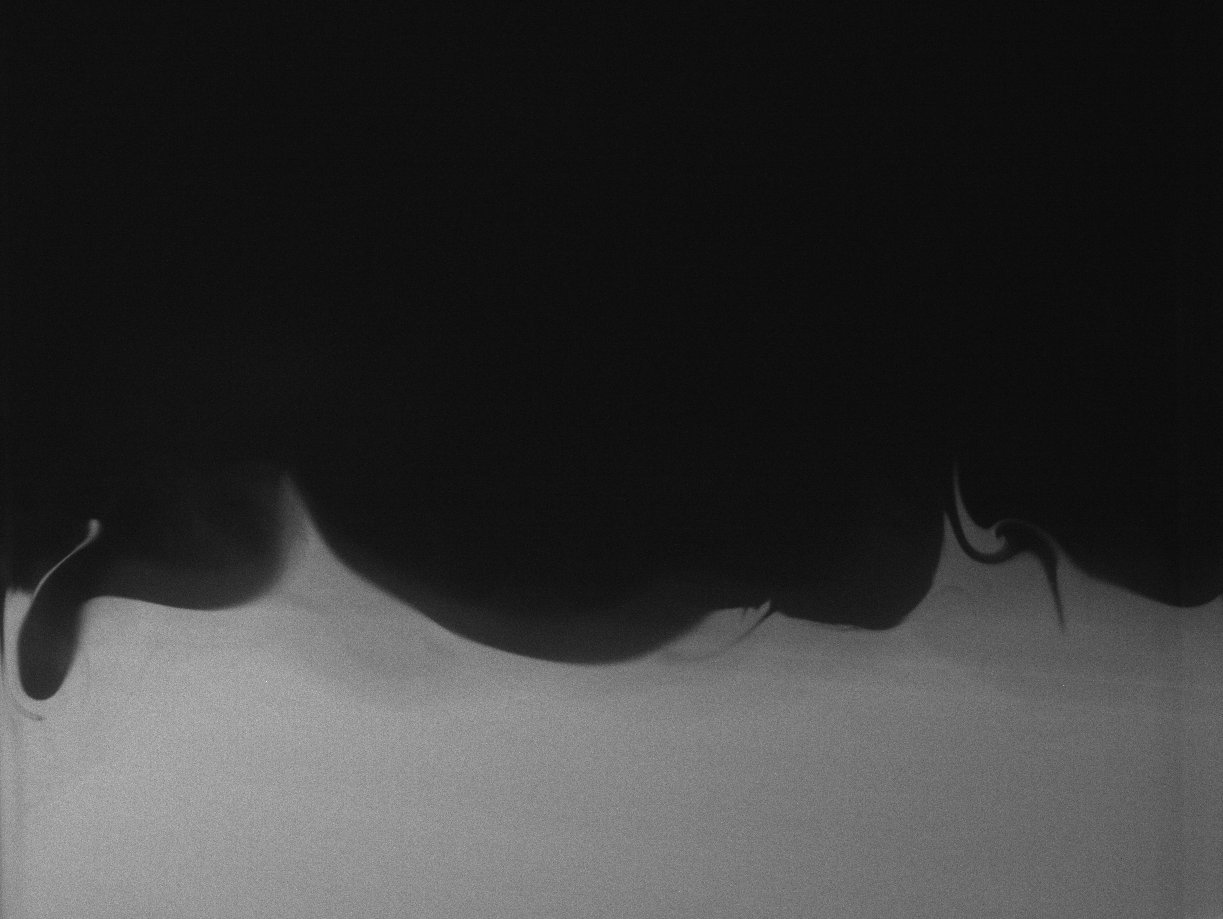}} \quad
	\subfloat[$t = 1440$ s.]{\includegraphics[width=0.35\textwidth]{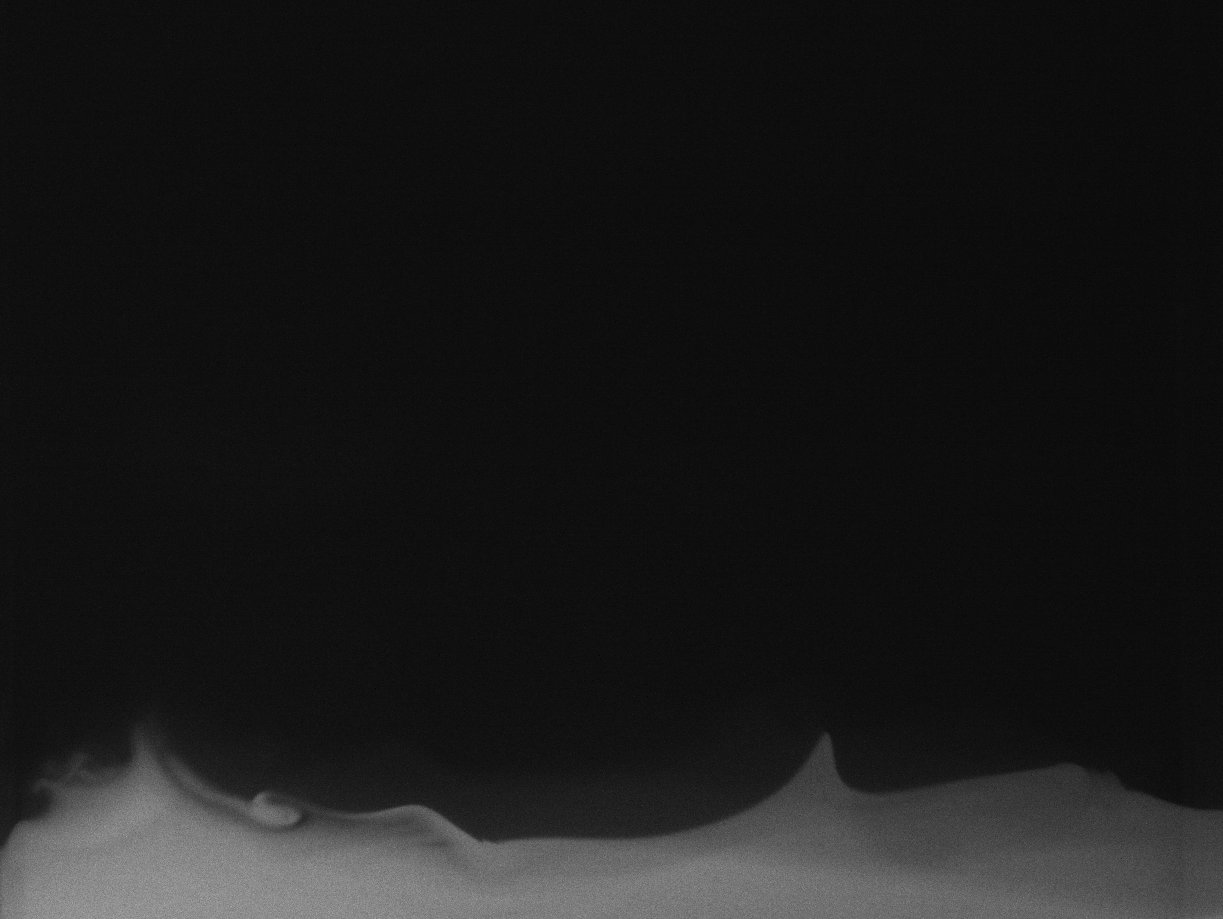}}
	\caption{Snapshots of LIF images at different times. The images are 20 cm wide and $t=0$ corresponds to the start of grid oscillations, whose lowest position is located 8 mm above the top of the images.}
	\label{fig:prop_regime}
\end{figure}
As shown in figure \ref{fig:prop_regime}, products and reactants remain separated by a sharp interface wrinkled by turbulence, but globally moving downward. The shape of the front, and thus the propagation velocity, depend on the local turbulence intensity, which decreases when going further away from the grid.
We now successively analyze the flow and the front velocities.

\subsubsection{Front propagation feedback on the flow}

\begin{figure}[H]
	\centering
	\includegraphics[width=\textwidth]{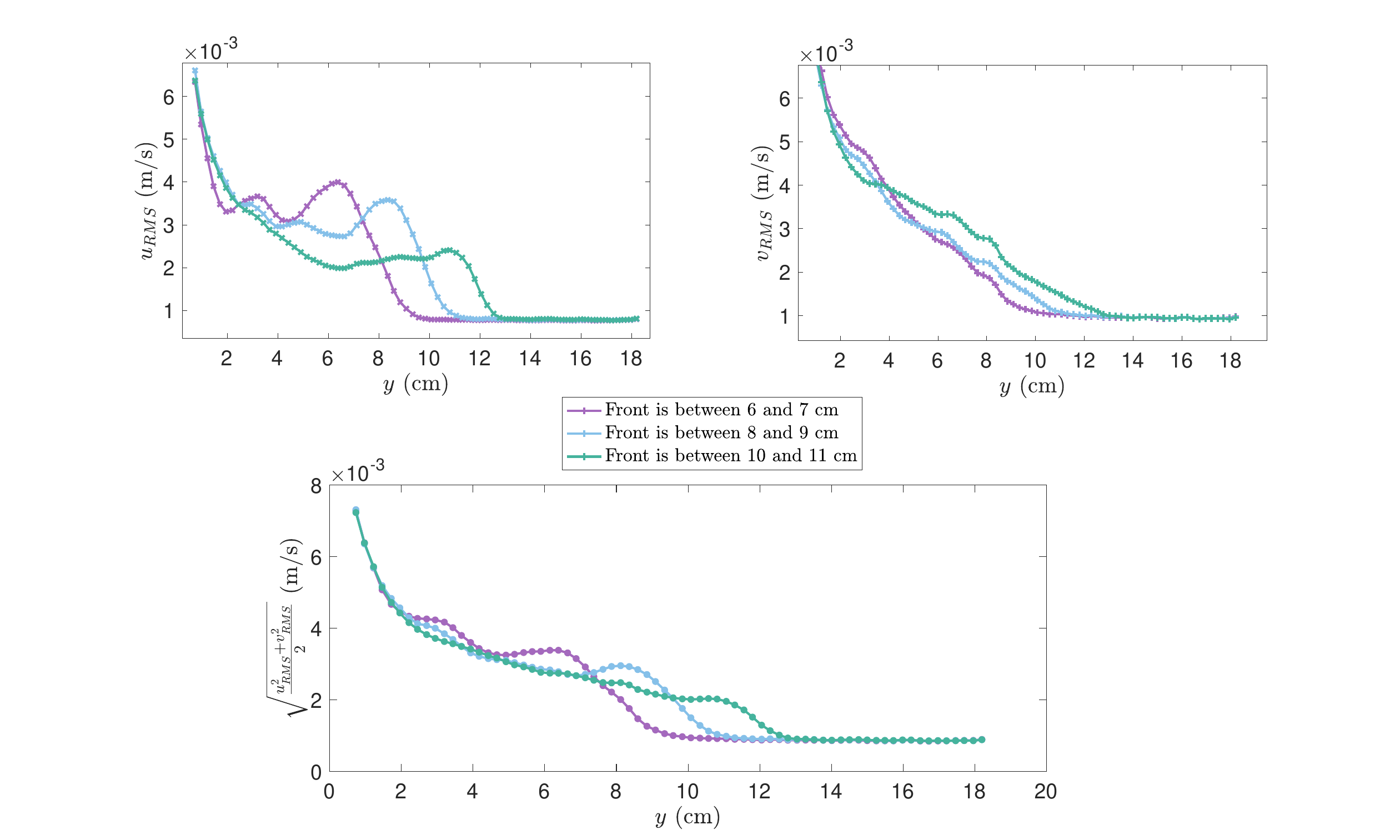}
	\caption{Flow short-time RMS velocity profiles of both components $u_{\text{RMS}}$ (horizontal) and $v_{\text{RMS}}$ (vertical) as well as their quadratic mean $\sqrt{(u_{\text{RMS}}^{2}+v_{\text{RMS}}^{2})/2}$ at different times, corresponding to different front positions.}
	\label{fig:u1g}
\end{figure}

Figure \ref{fig:u1g} shows the short-time RMS velocity profiles at different times, with
\begin{equation}
	u_{\text{RMS}}(y) = \overline{\sqrt{\left<u(x,y,t)^{2}\right>_{30 \text{s}}}} \quad \textrm{and} \quad
	v_{\text{RMS}}(y) = \overline{\sqrt{\left<v(x,y,t)^{2}\right>_{30 \text{s}}}} ,
\end{equation}
where $\Bar{\cdot}$ indicates the average over the x-direction and $<\cdot>_{30 \text{s}}$ indicates a time-average over 30 seconds. A front with a typical velocity of about 5 mm/min travels half a PIV box size in 30 seconds. Therefore, computing a RMS velocity over 30 seconds provides insight into the velocity field during a period over which the front has not significantly evolved vertically.

Clear feedback of the front propagation on the flow is observed in the horizontal velocity profiles in Figure \ref{fig:u1g}, with a significant overshoot of the horizontal velocity at the front, followed by a marked decrease below. The decrease in vertical velocity is less pronounced but still evident when compared to the expected dependence in the absence of a front (see previous section). A similar behavior was observed in a complementary experiment, using two layers of salty water with identical densities as the products and reactants (see supplemental material \cite{SM}). This confirms that the turbulence collapse at the interface is driven by buoyancy effects, not by the chemical reaction.

\begin{figure}[H]
	\centering
	\includegraphics[width=\textwidth]{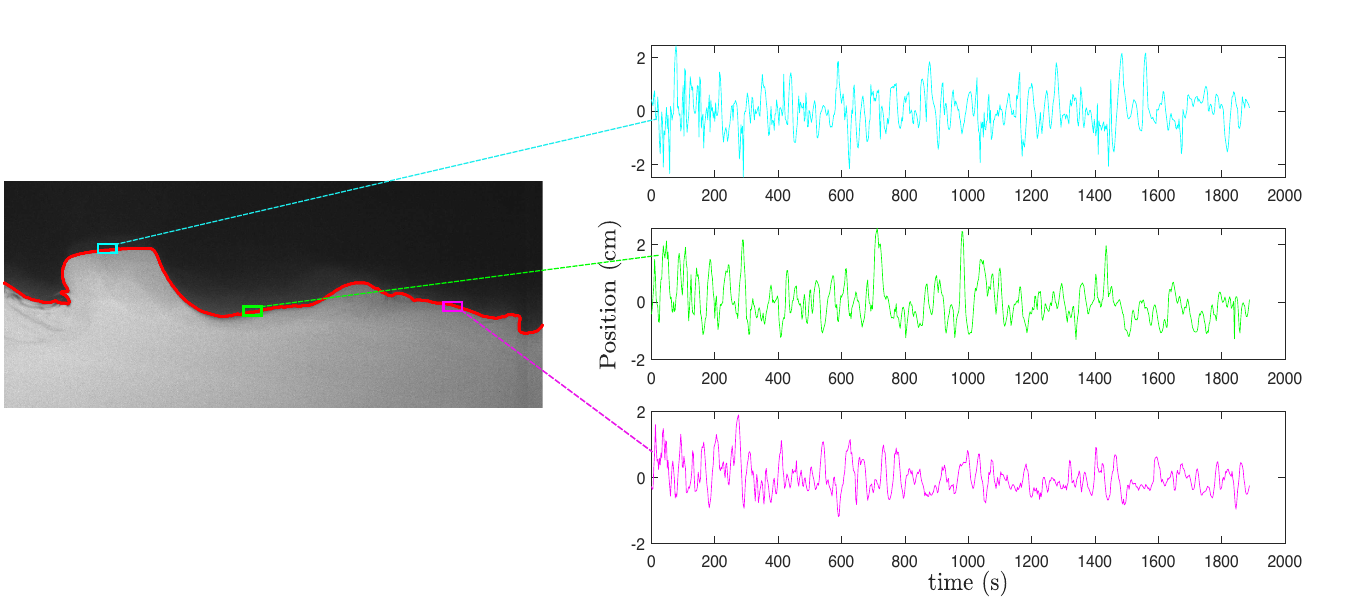}
	\caption{Time-evolution of position of parcels of the interface around the horizontal average front position. }
	\label{fig:OscillationInterface}
\end{figure}

\begin{figure}[H]
	\centering
	\includegraphics[width=0.5\textwidth]{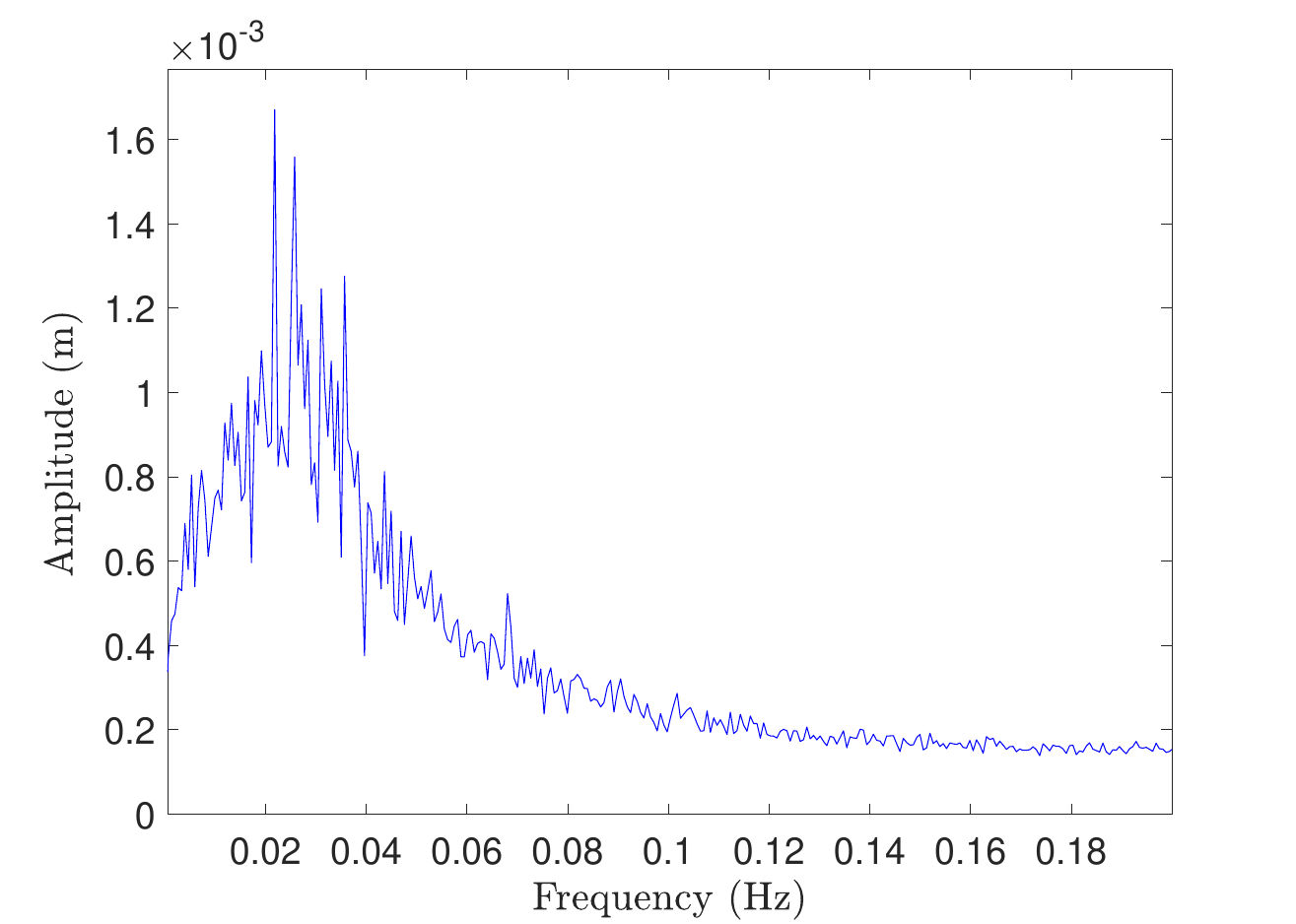}
	\caption{Temporal spectrum of the front oscillations, performed at each interface location and then averaged in the horizontal direction.}
	\label{fig:FTinterface}
\end{figure}    

We argue that actually, turbulent patterns impinging the interface transfer their energy to interfacial waves, explaining both the emergence of anisotropy with the amplitude increase of the horizontal velocity at the front, and the velocity decrease below the front. To further illustrate this, we use LIF measurements and plot in figure \ref{fig:OscillationInterface}  three examples of the temporal front fluctuations around its horizontal mean at three different abscissa (average over 40 pixels $\approx$ 6 mm). The three curves display a similar tendency: a slightly decreasing amplitude due to the front getting further from the grid along with what seems to be a similar predominant period. To retrieve this period, a Fourier transform is applied to the time evolution of the fluctuations around the horizontal mean at all points on the front, following a high-pass filter with a cut-off frequency of $f_{\text{cut-off}}=0.005$ Hz. After averaging all the resulting Fourier transforms, figure \ref{fig:FTinterface} is obtained. The spectrum shows a peak at $f \sim 0.025 $ Hz, corresponding to a period of $T \sim 1/0.025 = 40$s. In a two-layer fluid and in the limit of deep water, the dispersion relation of interfacial waves is 
\begin{equation}
	\omega =  \sqrt{\frac{\Delta \rho}{\rho}g\frac{\pi}{\lambda}}
\end{equation}
where $\omega = 2\pi f$ is the angular frequency and 
$\lambda$ the wavelength, hence giving a period $T$ 
\begin{equation}
	T = \frac{2\pi}{\sqrt{\frac{\Delta \rho}{\rho}g\frac{\pi}{\lambda}}}.
\end{equation}
Taking orders of magnitude for the density difference  $\frac{\Delta \rho}{\rho} \approx 10^{-4}$ and an estimated wavelength (peak-to-peak distance) of order $\lambda \approx \frac{H}{2} = 0.1$~m (cf. figure \ref{fig:OscillationInterface}), we obtain $T \approx 36$~s, a value close to the one found on figure \ref{fig:FTinterface}, supporting the fact that the oscillation frequency of the interface is due to the presence of gravity waves.

\subsubsection{Turbulent front propagation velocity}

To compute the front velocity, LIF images are binarized to assign the value 0 to the pixels where products are located (black) and 1 to the reactants (white). In this case, the horizontal-mean vertical location of the reactants at the $n^{\text{th}}$ image corresponds to
\begin{equation}
	Z_{f}(t=t_{n}) = \frac{\mathcal{C}}{N_{x}}\sum_{j=1}^{N_{x}}\sum_{k=1}^{N_{y}}\mathcal{I}_{n}(j,k),
	\label{eq:AvFrontPos}
\end{equation}
where $N_{x}$ is the number of pixels in the horizontal direction, 
$N_{y}$ is the number of pixels in the vertical direction, 
$\mathcal{I}_{n}$ a matrix of size $N_{x} \times N_{y}$ containing the pixel values of the binarized n$^{\text{th}}$ image (1 for reactants and 0 for products) and $\mathcal{C}$ the calibration factor in m/pixel. The mean, turbulent front velocity is then given by
\begin{equation}
	S_{T}(t) = \abs{\frac{dZ_{f}(t)}{dt}}.
\end{equation}

\begin{figure}[H]
	\centering
	\subfloat[Stretching]{\includegraphics[width=0.25\textwidth]{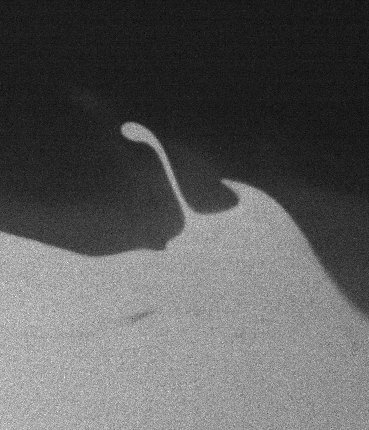}}
	\subfloat[Detachment]{\includegraphics[width=0.25\textwidth]{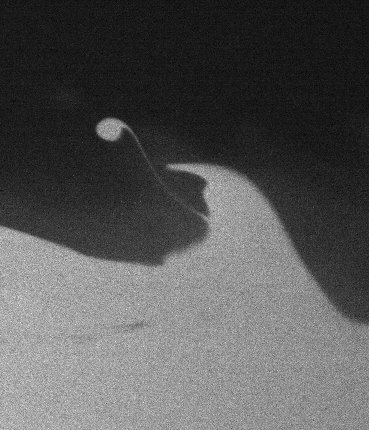}}
	\subfloat[Transport]{\includegraphics[width=0.25\textwidth]{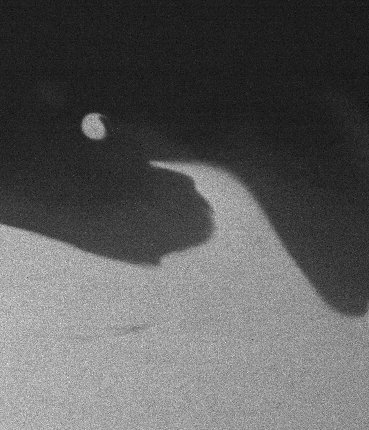}}
	\caption{Example of a blob of reactants detachment from the front and its advection inside the products in the case of an oscillating grid located in the products. Zoom on a 6 cm $\times$ 7 cm area around a portion of the interface.}
	\label{fig:filament_reactants}
\end{figure}

Besides, in order to quantify the local turbulence intensity at the front, the following steps are implemented at each time:
\begin{enumerate}
	\item detect the interface, 
	\item delimit a zone centered around the interface, of 64 px width  (i.e. two PIV boxes),
	\item project the resulting band on the 16 $\times$ 16 px$^{2}$ PIV grid (we use PIV  boxes of size $32 \times 32$ px$^{2}$ and an overlap of 50\%),
	\item find the corresponding PIV velocities $u$ and $v$ contained inside this band,
	
	\item compute the root mean square of all the velocities contained in the band 
	\[
	\begin{cases}
		u_{f} = \sqrt{\dfrac{\sum u^{2}}{N_{pts}}} \\
		v_{f} = \sqrt{\dfrac{\sum v^{2}}{N_{pts}}} \\
		u_{\text{RMS}} = \sqrt{\frac{u_{f}^{2}+v_{f}^{2}}{2}}
	\end{cases}
	\]
	where $N_{pts}$ is the number of velocity points retrieved inside the band. 
\end{enumerate}
Note that when filaments of reactants detach from the interface and get advected towards the oscillating grid (see figure \ref{fig:filament_reactants}), they are not taken into account in this velocity measurement. Their volume, hence their contribution, however remain negligible in the present regime.

\begin{figure}[H]
	\centering
	\includegraphics[width=\linewidth]{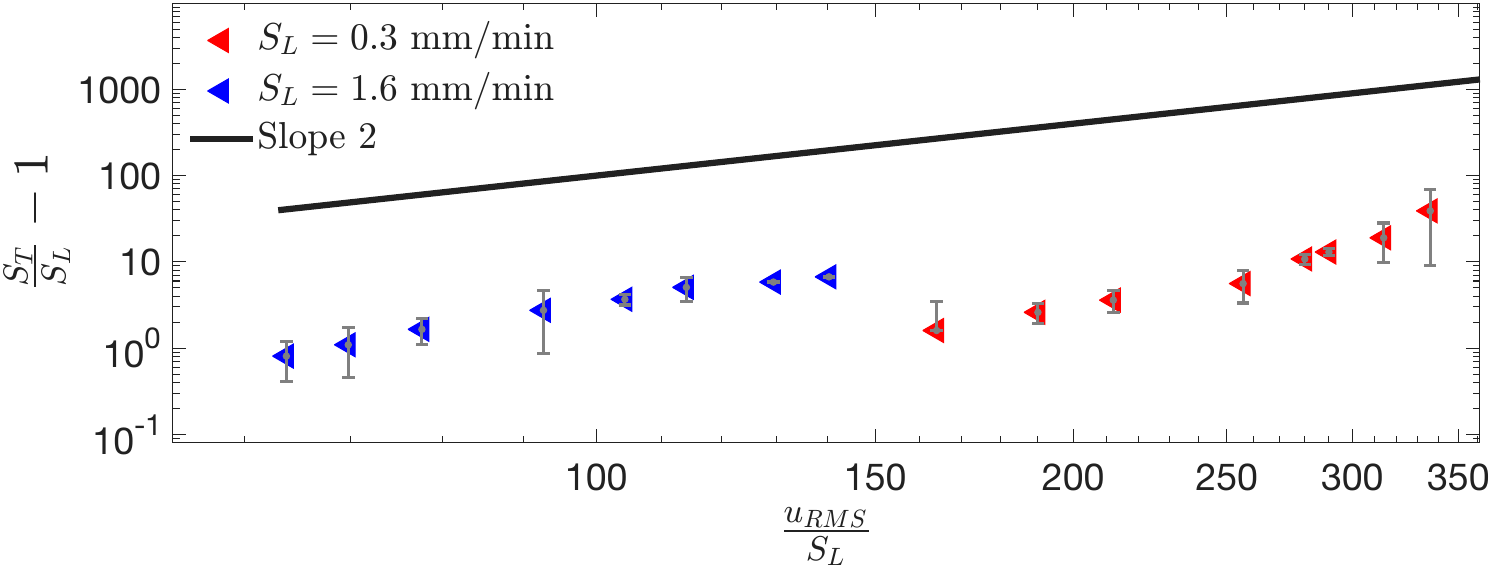}
	\caption{Front velocity as a function of the local turbulence intensity in spatially decaying turbulence when the grid is oscillating at the top of the tank. Blue/red triangles correspond to SOL1/SOL2, respectively. Error bars show velocity uncertainty propagated from the interface’s spatial spread.}
	\label{fig:1glaw}
\end{figure}

Figure \ref{fig:1glaw} shows the evolution of the turbulent front velocity $S_T$ as a function of the RMS flow velocity $u_{\text{RMS}}$ for both {SOL1} and {SOL2}. Both velocities are made dimensionless by the laminar front velocity $S_L$, and since we are interested in the turbulent acceleration of the front, we actually plot $S_T/S_L-1$.
The best fit using a power function gives an exponent $ 1.9\pm 0.1$ and a prefactor of $6.7\cdot 10^{-4} \pm 2.0 \cdot 10^{-4}$ for the experiment at $S_{L} = 1.6$ mm/min and an exponent of $ 3.1\pm 0.3$ and a prefactor of $4.4\cdot 10^{-7} \pm 3.5 \cdot 10^{-7}$ for the experiment at $S_{L} = 0.3$ mm/min. The gap between the two point series indicates the possible influence of parameters other than $u_{\text{RMS}}$ and $S_{L}$ on $S_{T}$, which will be addressed in section \ref{section:2g}. However, the quadratic scaling law in equation (\ref{eq:CW79})  agrees well with experimental results, even for large interface deformations at high velocity ratios $u_{\text{RMS}}/S_L$.

\subsection{Reactive mixing regime}

In the Boussinesq approximation, the flow produced by either the top or bottom grid is rigorously symmetrical, even in the presence of a density interface, as verified in our complementary salty water experiment with no chemical reaction but identical density effects (see supplemental material \cite{SM}). Placing the energy source at the bottom in the denser fluid or at the top in the less dense fluid generates the same turbulence and intensity decay, provided the source is equidistant from the interface. However, this symmetry breaks when considering the reaction, due to its inherent directionality: the reaction evolves from products to reactants, i.e., downward in our case.

\begin{figure}[H]
	\centering
	\subfloat[$t= 43$s.]{\includegraphics[width=0.25\textwidth]{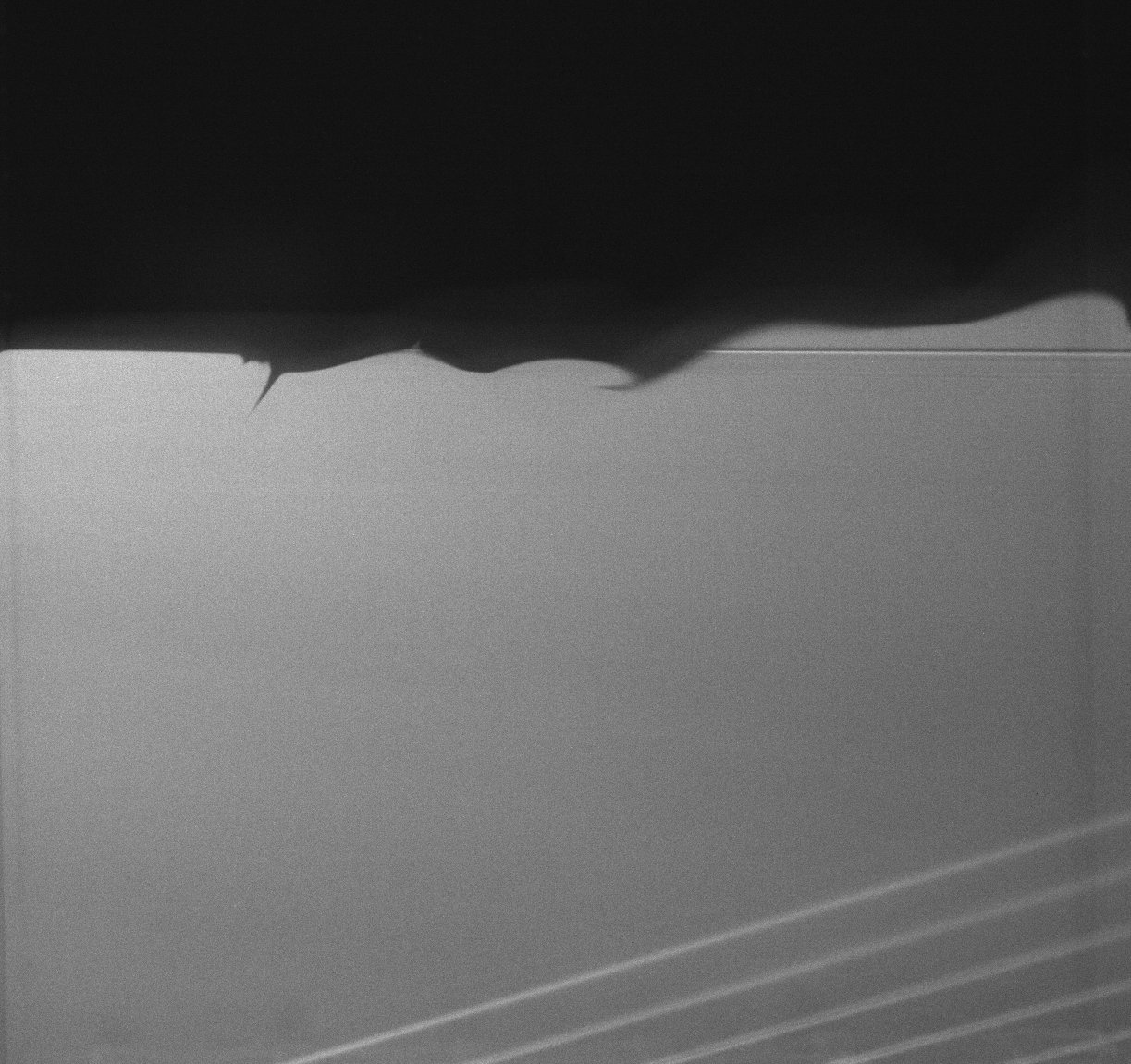}} 
	\subfloat[$t= 198$ s.]{\includegraphics[width=0.25\textwidth]{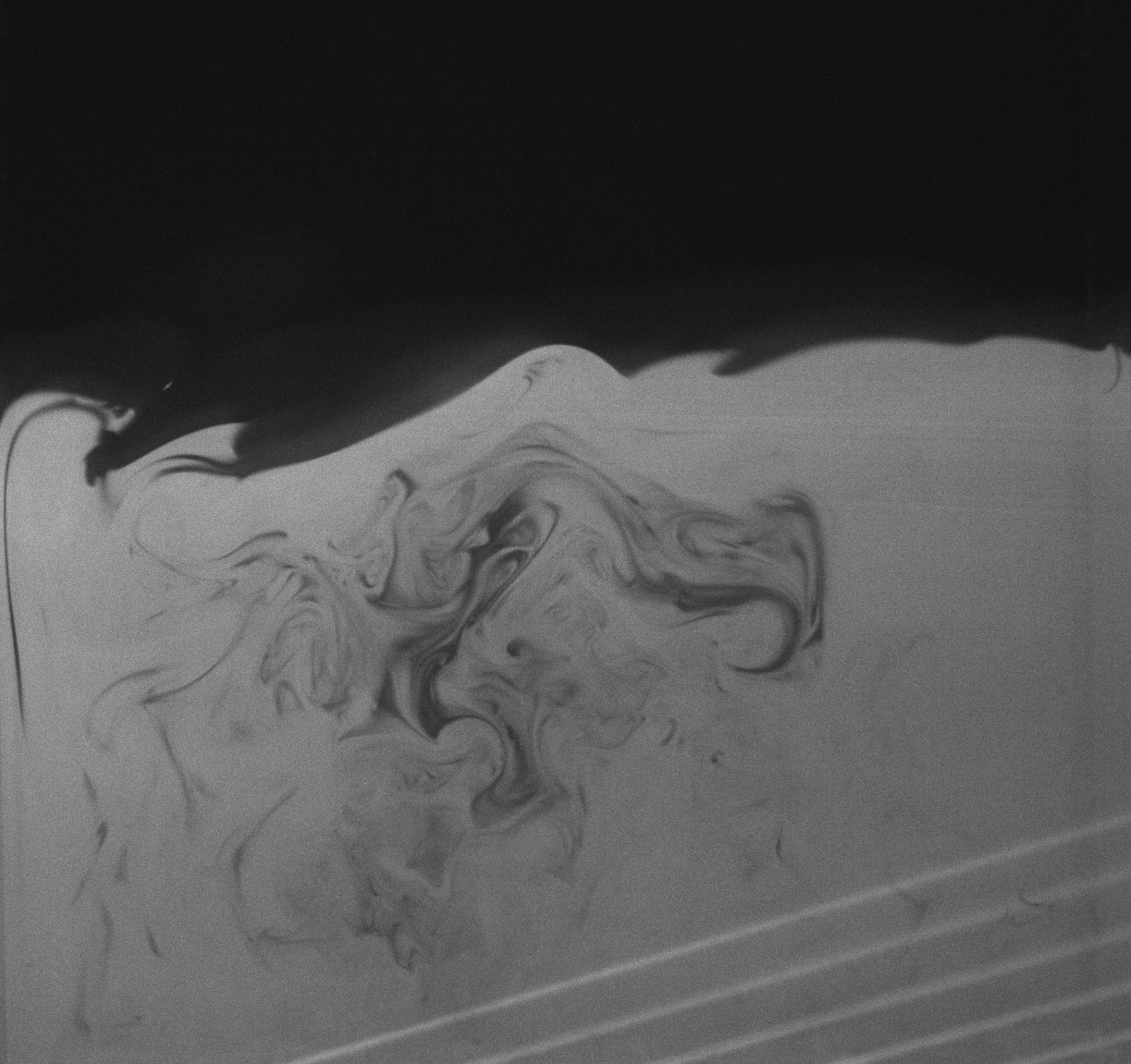}}
	\subfloat[$t=390$ s.]{\includegraphics[width=0.25\textwidth]{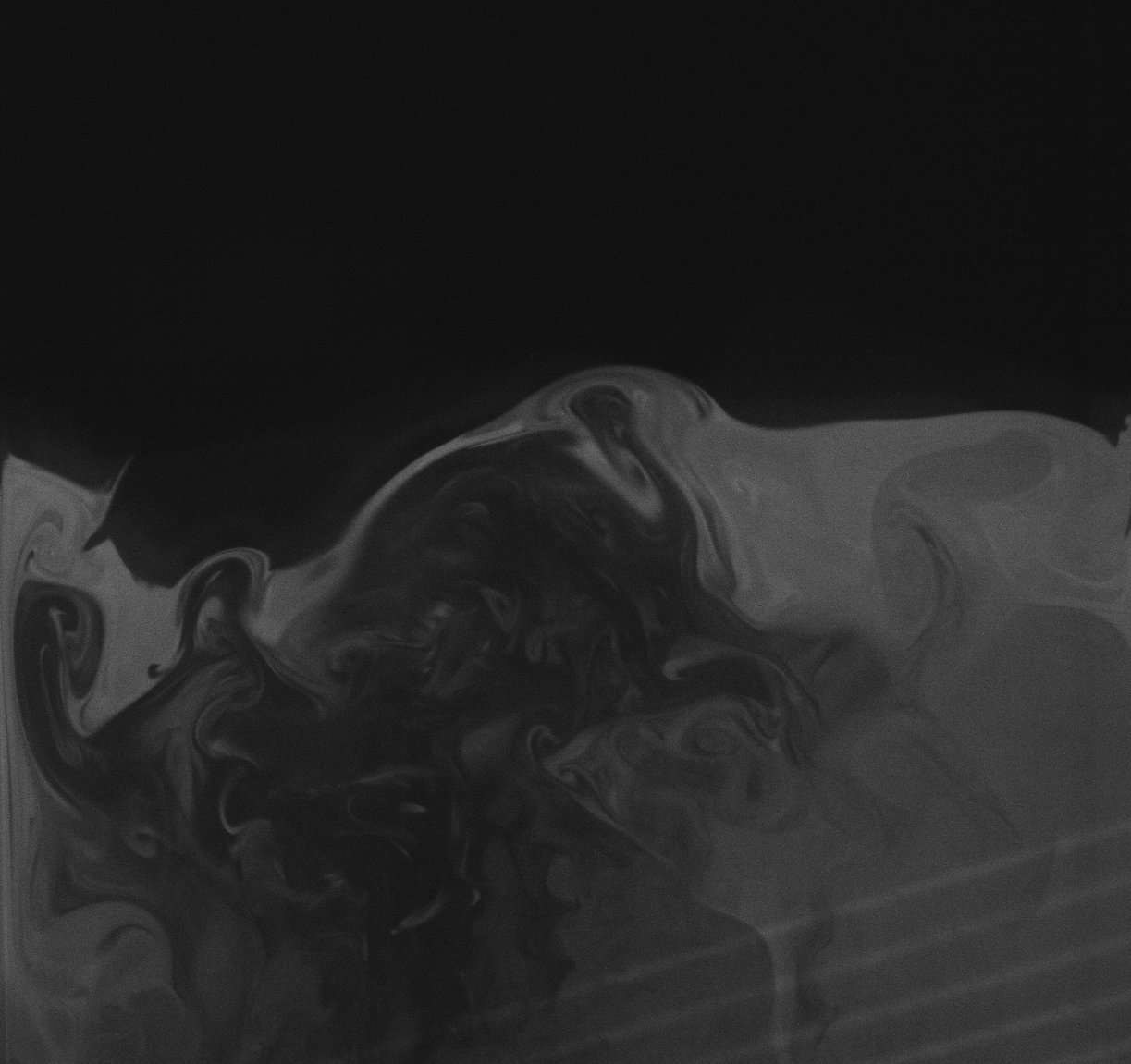}}
	\subfloat[$t = 483$ s.]{\includegraphics[width=0.25\textwidth]{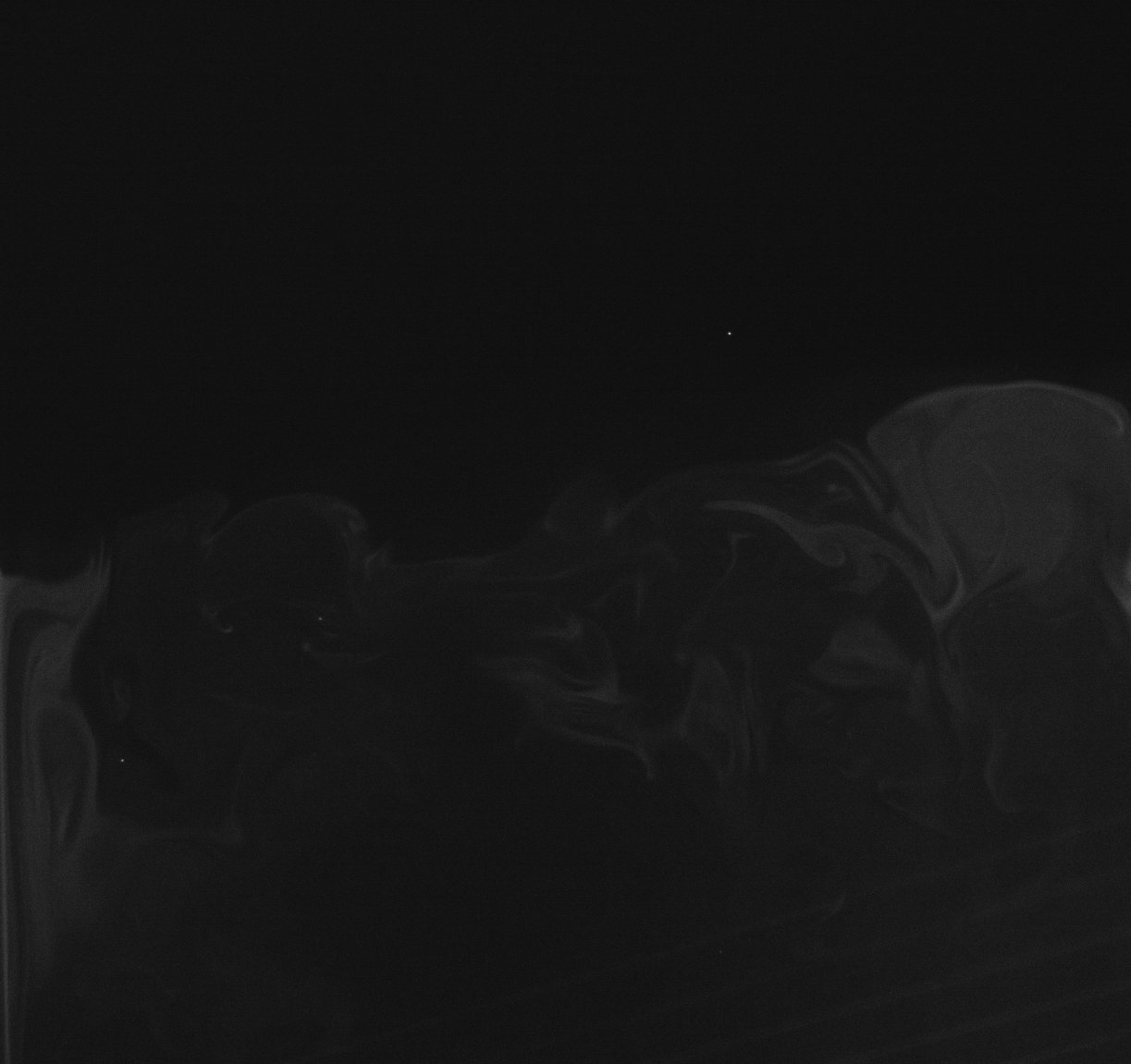}}
	\caption{Snapshots of LIF images for {SOL1} at different times. The images are 20 cm wide and $t=0$ corresponds to the start of grid oscillations positioned at the bottom of the tank (in the reactants), with the highest position located $\approx 1$ cm below the bottom of the images.}
	\label{fig:mixing_regime}
\end{figure}

In section \ref{section:propagationregime}, turbulence was generated inside the products, causing the front to wrinkle while remaining sharp, with a continuous material interface separating the products and reactants. We now investigate what happens when the grid oscillates only in the reactants, at the bottom of the tank. As before, the tank is filled with the solution and after a laminar propagation phase, the oscillations are initiated at $f = 6$ Hz and $A = 10$ mm. Typical evolution is shown in figure \ref{fig:mixing_regime}.

\begin{figure}[H]
	\centering
	\subfloat[Stretching]{\includegraphics[width=0.25\textwidth]{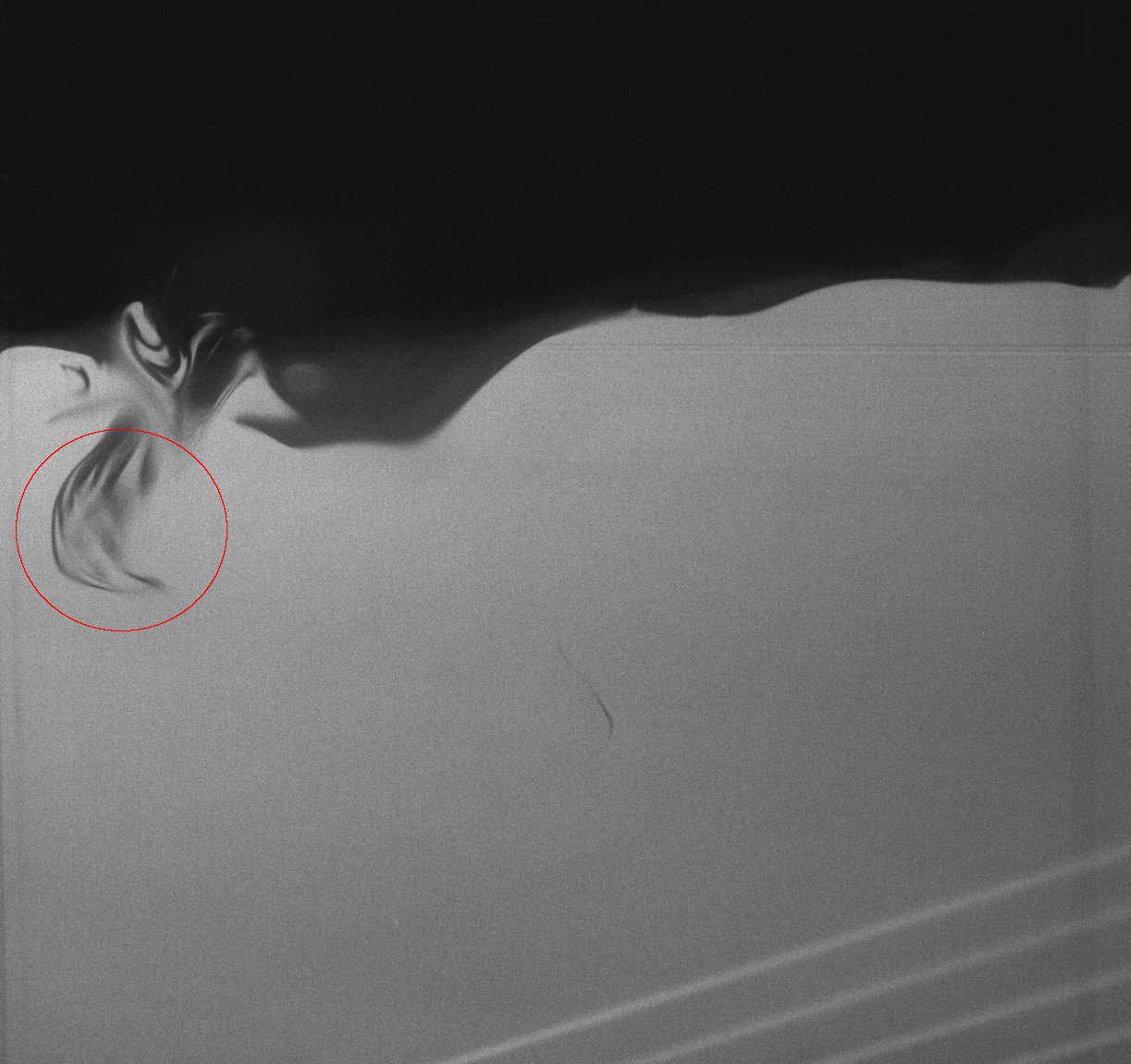}}
	\subfloat[Detachment]{\includegraphics[width=0.25\textwidth]{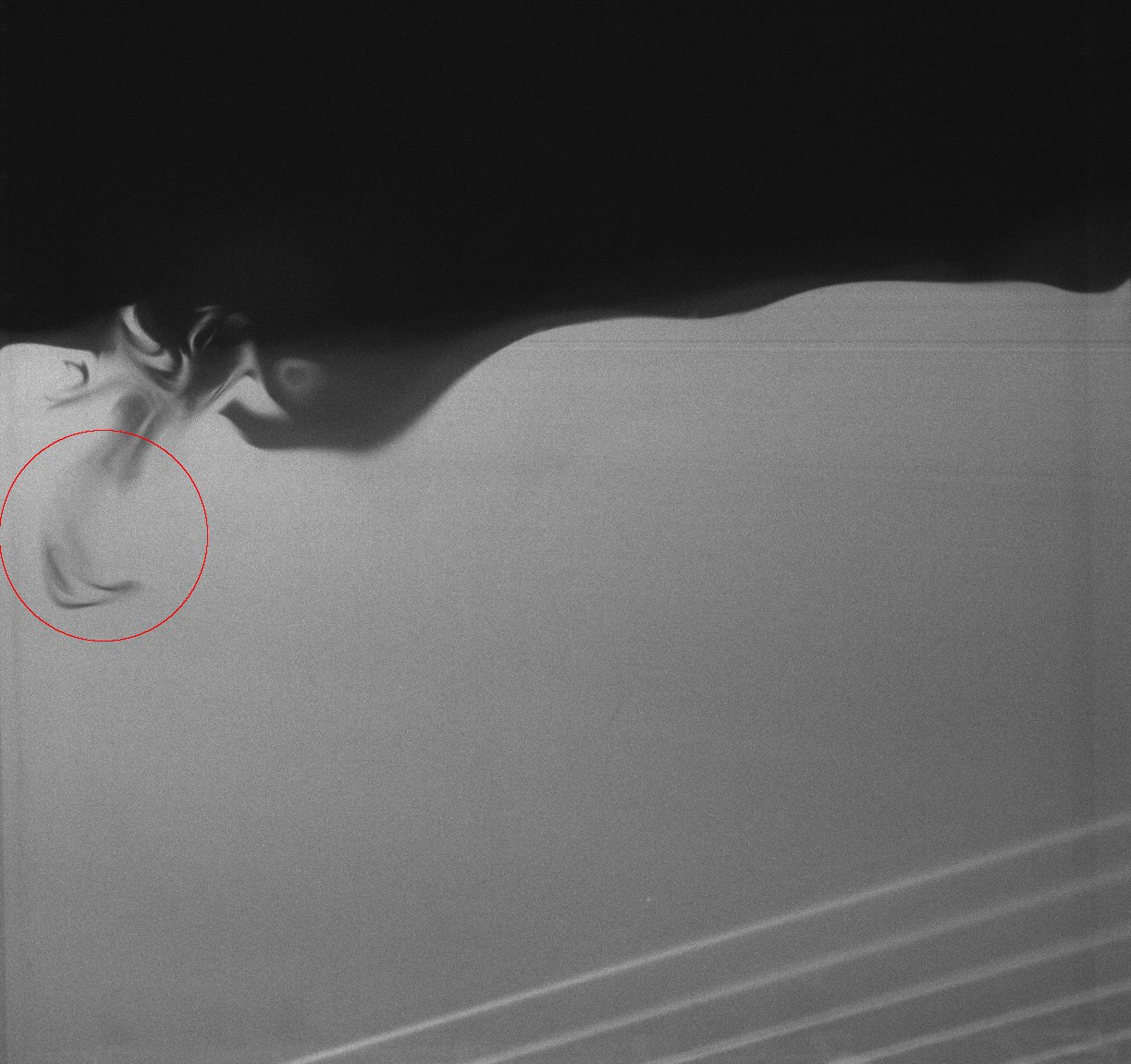}}
	\subfloat[Transport ]{\includegraphics[width=0.25\textwidth]{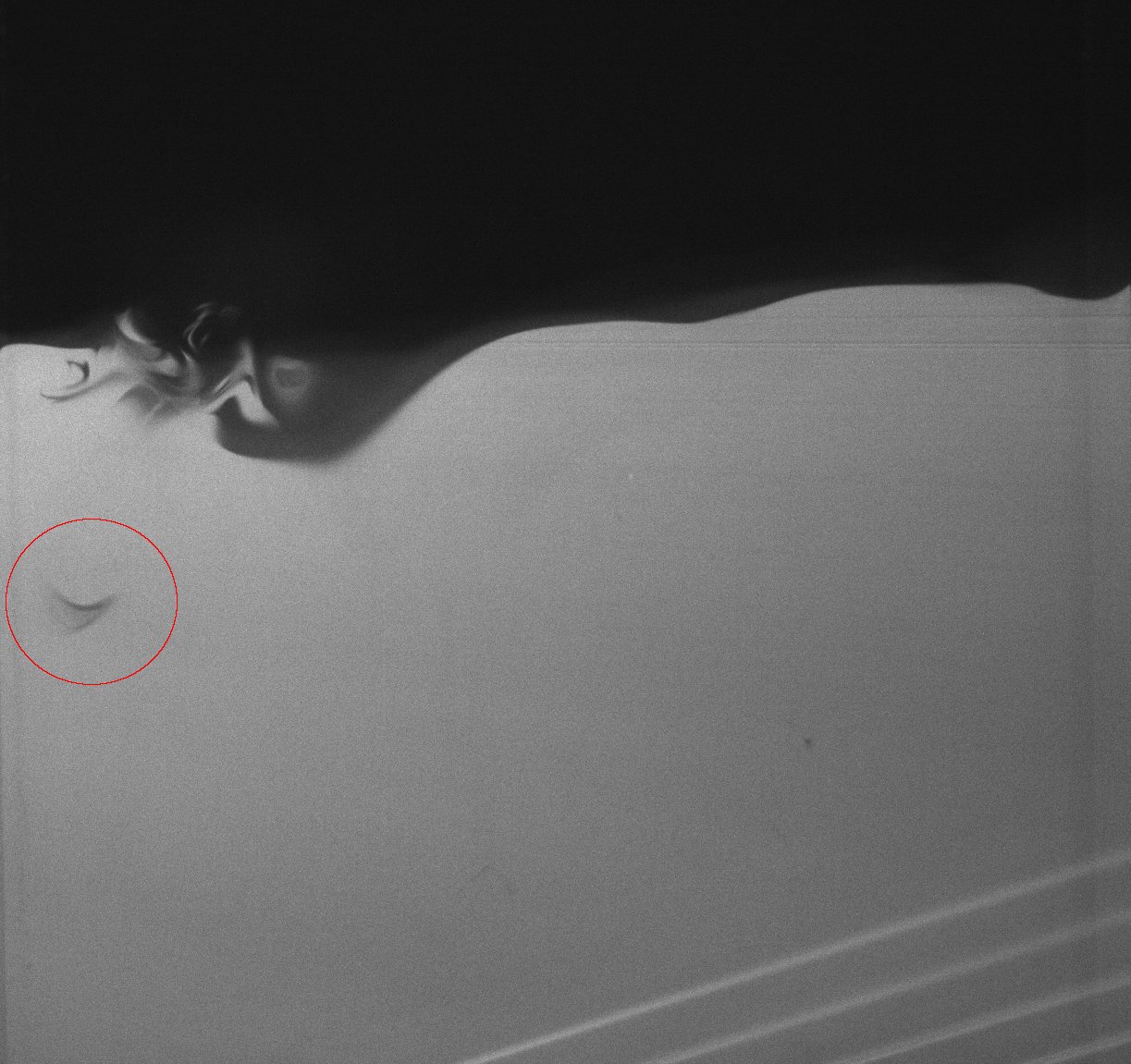}}
	\caption{Example of a blob of products detachment from the front and its advection inside the reactants in the case of an oscillating grid located in the reactants.}
	\label{fig:filament_products}
\end{figure}
\vspace{-0.5cm}
For the first $\sim 80$ seconds, the front stretches as usual, with cusps forming locally along it. However, as illustrated in figure \ref{fig:filament_products}, some product blobs are advected into the reactant bulk towards the bottom oscillating grid. In the previous section, we saw that a filament of reactants transported inside the products reacts instantaneously without affecting the front dynamics (figure \ref{fig:filament_reactants}).
When the grid is located in the reactants, each time a catalyst-rich lamella of products detaches from the front and gets pulled into the reactants, it locally initiates the reaction. This causes the reaction to begin at different points in the bulk, particularly near the grid, where the flow velocity (and thus mixing) is large. Consequently, the reaction proceeds rapidly and quasi-homogeneously in the bulk, as shown by the global intensity decrease in figure \ref{fig:mixing_regime}, until the interface disappears. In this case, the Huygens' model is irrelevant. As described in \cite{ou1983}, mixing entrains and stretches product lamellae containing the catalyst  into the reactant region, where the catalyst diffuses into the surrounding reactants and locally ignites the reaction. The product distribution in this case depends on the reaction rate, reactant size, and mixing conditions.
To distinguish this newly identified regime from the one where the interface remains continuous and the classical Huygens' model applies, we refer to it as the {reactive mixing regime}. Since it lies outside the scope of our main focus, it will not be further analyzed here. However, the effect of the bottom grid should be kept in mind for the upcoming investigation using two oscillating grids to produce isotropic turbulence.
\vspace{-0.2cm}
\section{Front propagation in quasi homogeneous and isotropic turbulence}
\label{section:2g}

\begin{figure}[H]
	\centering
	\subfloat[]{\includegraphics[width=0.25\textwidth]{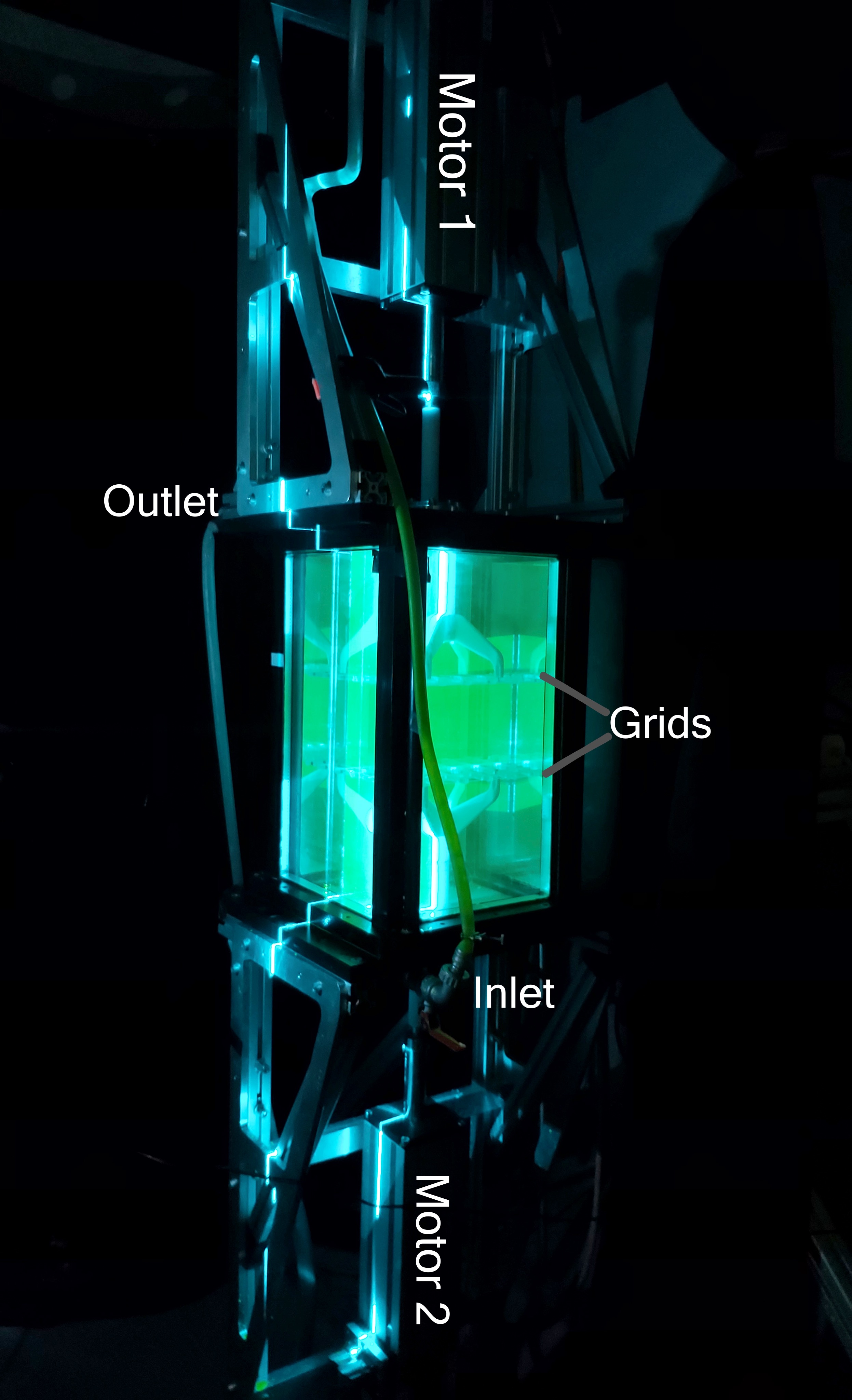}} \quad
	\subfloat[]{\vspace{1.2cm}\includegraphics[width=0.325\textwidth]{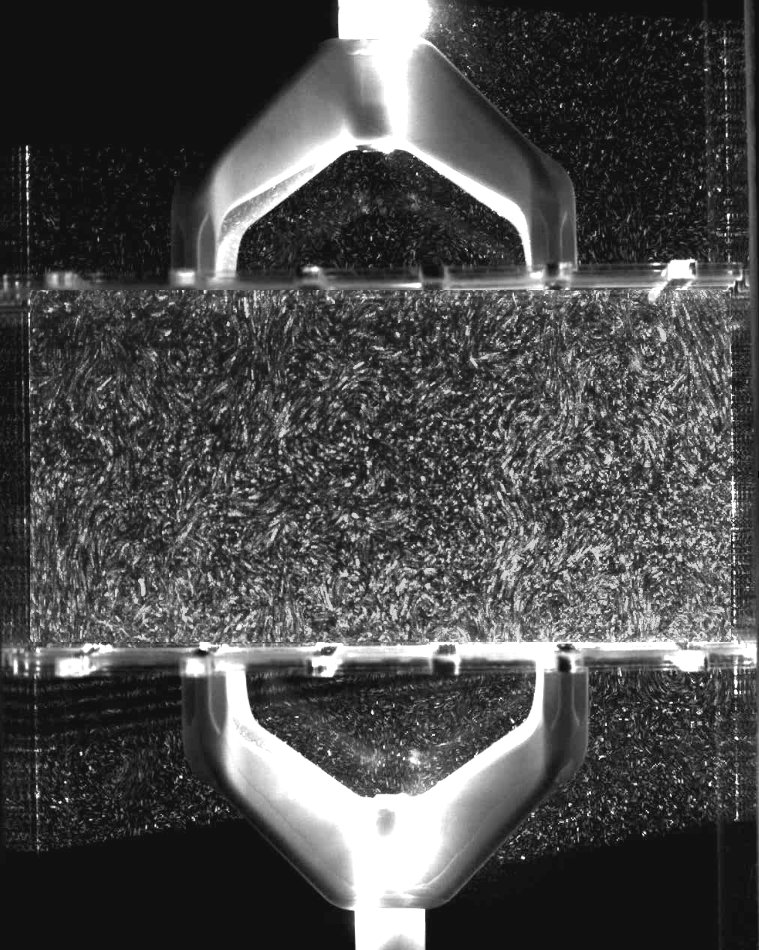}}
	\caption{Experimental apparatus with 2 grids. (a) Global view of the set-up and (b) front view of the working area with PIV particles and a long time exposure of 1s. }
	\label{fig:PhotoManip2}
\end{figure}

We now want to get rid of the vertical flow inhomogeneity arising in a single grid set-up and study front propagation in an homogeneous isotropic turbulence. Many experimental facilities have been  designed for this purpose by the mean of different actuators, such as fans \cite{birouk2003}, pumps \cite{mccutchan2023}, magnetic fields \cite{gorce2022}, rotating disks \cite{verhille2021}, etc. Here we use a system of two oscillating grids, as  introduced by \cite{villermaux1995}. Yet, rather than using a single stirrer for the two grids connected with a rod, we use two distinct motors oscillating in phase opposition, as shown in figure \ref{fig:PhotoManip2}. 

\subsection{Flow characterization}

Our first goal is to validate the homogeneity and isotropy of the flow, and to establish a scaling law for the turbulence intensity as a function of the different control parameters, expending the law \eqref{eq:loi_hopfinger} to also include the influence of the gap between the two grids $G$. 
While the conditions listed in \ref{section:geometricalrequirements} to limit the presence of large-scale recirculation in the one-grid case remain mandatory, an additional requirement must be met when working with two grids:  
since the individual jets produced by each grid are detectable at a distance of up to $M$ [13], a sufficient gap between both grids is necessary for the jets to no longer be discernible. A minimum gap of 100 mm was found to be sufficient in our case.
\label{eq:conditionG}

\begin{figure}[H]
	\centering
	\includegraphics[width=0.8\textwidth]{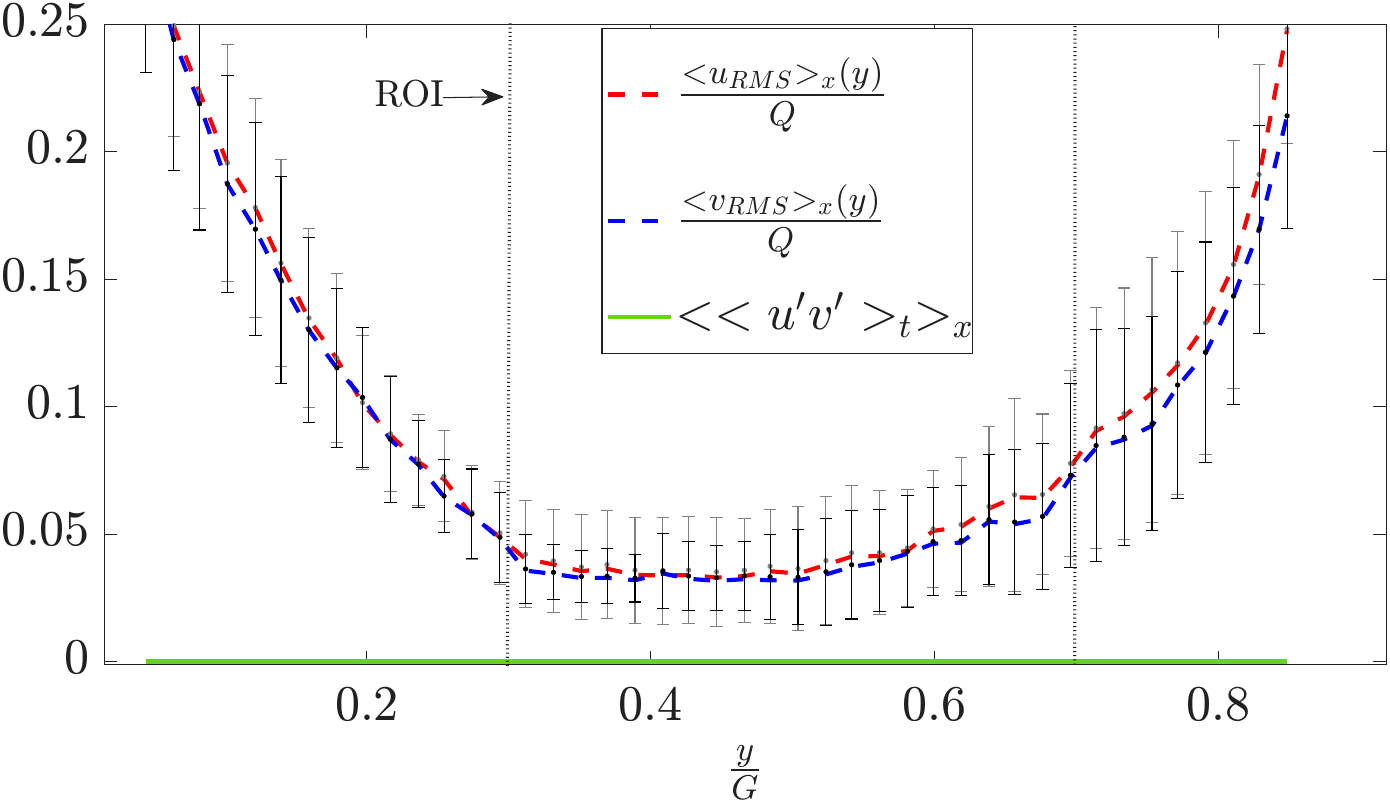}
	\caption{Evolution of the RMS velocities $u_{RMS}$ and $v_{RMS}$ normalized by the grid speed $Q = fA$ and of the Reynolds stress tensor $<u'v'>$ normalised by $Q^2$ as a function of the distance along the oscillation axis $\Vec{e_{y}}$, with the two grids oscillating around $y/G=0$ and $y/G=1$ respectively, for $G = 140$ mm, $f = 3$ Hz, $A = 10$ mm. In all cases, the results are averaged in the $x$-direction. The region of interest (ROI) for our measurements is located between $y/G=0.3$ and $y/G=0.7$. Error bars show the standard deviation across the $\Vec{x}$-direction, in grey for $u_{RMS}$ and in black for $v_{RMS}$.}
	\label{fig:PlateauIsotropie}
\end{figure}

In this section, experiments are performed in fresh water. 
Figure \ref{fig:PlateauIsotropie} represents the evolution along the $\Vec{e_{y}}$ axis of the RMS velocities. 
Since each grid is a source of  turbulence, the maximum RMS velocity is reached in their vicinity, then decreases following a power law as for a single grid. But around the center of the two-grid system, the cumulative turbulent intensity reaches a plateau where $u_{RMS}(y)$ and $v_{RMS}(y)$ remain quasi constant and of similar value, indicating the region where turbulence is nearly homogeneous and isotropic. 
In this region, the degree of isotropy $\dfrac{u_{\text{RMS}}}{v_{\text{RMS}}}$ is around 0.976, which is better than in the case of a stationary wind tunnel grid ($\approx 1.3$) \cite{corrsin1963}. One can further notice that the Reynolds stress is found to be essentially zero. Furthermore, when moving the laser sheet to different plans perpendicular to the grids while staying at least at a minimum distance $M/2$ from the walls, similar values of turbulent intensities are found.

\begin{figure}[H]
	\hspace{-2cm}
	\includegraphics[width=1.2\textwidth]{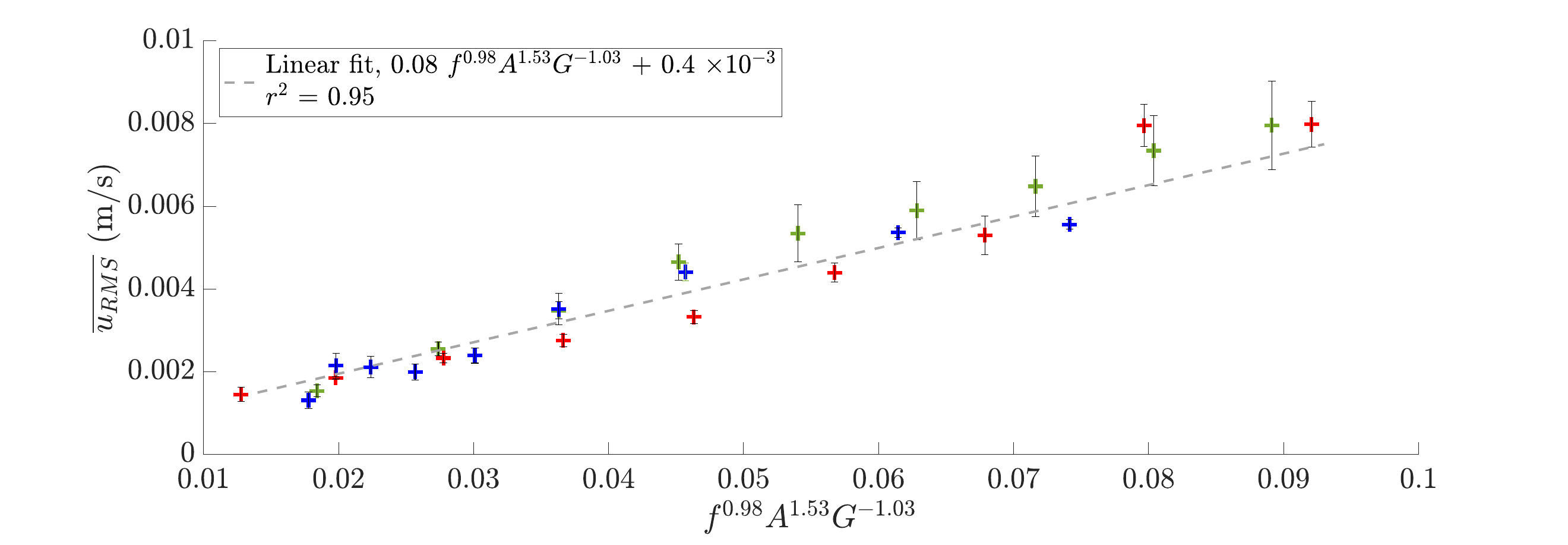}
	\caption{Evolution of $\overline{u_{RMS}}$ where $\Bar{.}$ denotes the spatial average over the whole ROI. The dashed line corresponds to law \eqref{eq:loi_globale}, determined from studying separately the influence of the gap (blue markers), the frequency (green) and the amplitude (red). Here, only the points in the range of interest regarding the Reynolds number and the conditions to avoid mean-flow production are taken into account. The error bars indicate the standard deviation over the region of interest and the exponents on $f$, $A$ and $G$ (respectively 0.98, 1.53 and -1.03) are determined from best fits to the results obtained from systematic variations of each parameter separately.} 
	\label{fig:loi_globale}
\end{figure}

To determine the influence of each of the control parameters $f,A,G$ on the turbulent intensity in the region of interest (ROI), a systematic study is carried out where each parameter is individually varied while the rest of the parameters is kept constant. We set (i) $G = 100$ mm, $A= 10$ mm, and frequencies $f$ ranging from 1 to 10 Hz, (ii) $G = 100$ mm,  $f= 2$ Hz, and amplitudes $A$ ranging from 5 to 50 mm, (iii) $A = 10$ mm, $f= 4$ Hz, and gaps $G$ ranging from 50 to 200 mm.
From there, as shown on figure \ref{fig:loi_globale}, the following scaling law describing the turbulent intensity was established 
\begin{equation}
	u_{RMS} = k fA^{1.5}G^{-1},
	\label{eq:loi_globale}
\end{equation}
where $k=0.08$ m$^{\frac{1}{2}}$. This is in good agreement with  Hopfinger \textit{et al.} (1975) \cite{hopfinger1975}, who further found that the dependence on the mesh size is in $M^{\frac{1}{2}}$, in agreement with the one-grid law (\ref{eq:loi_hopfinger}). 

\subsection{From the propagation to the reactive mixing regime}
\label{section:frompropagtomix}
\begin{figure}[H]
	\centering
	\subfloat[t = 0.]{\includegraphics[width=0.25\textwidth]{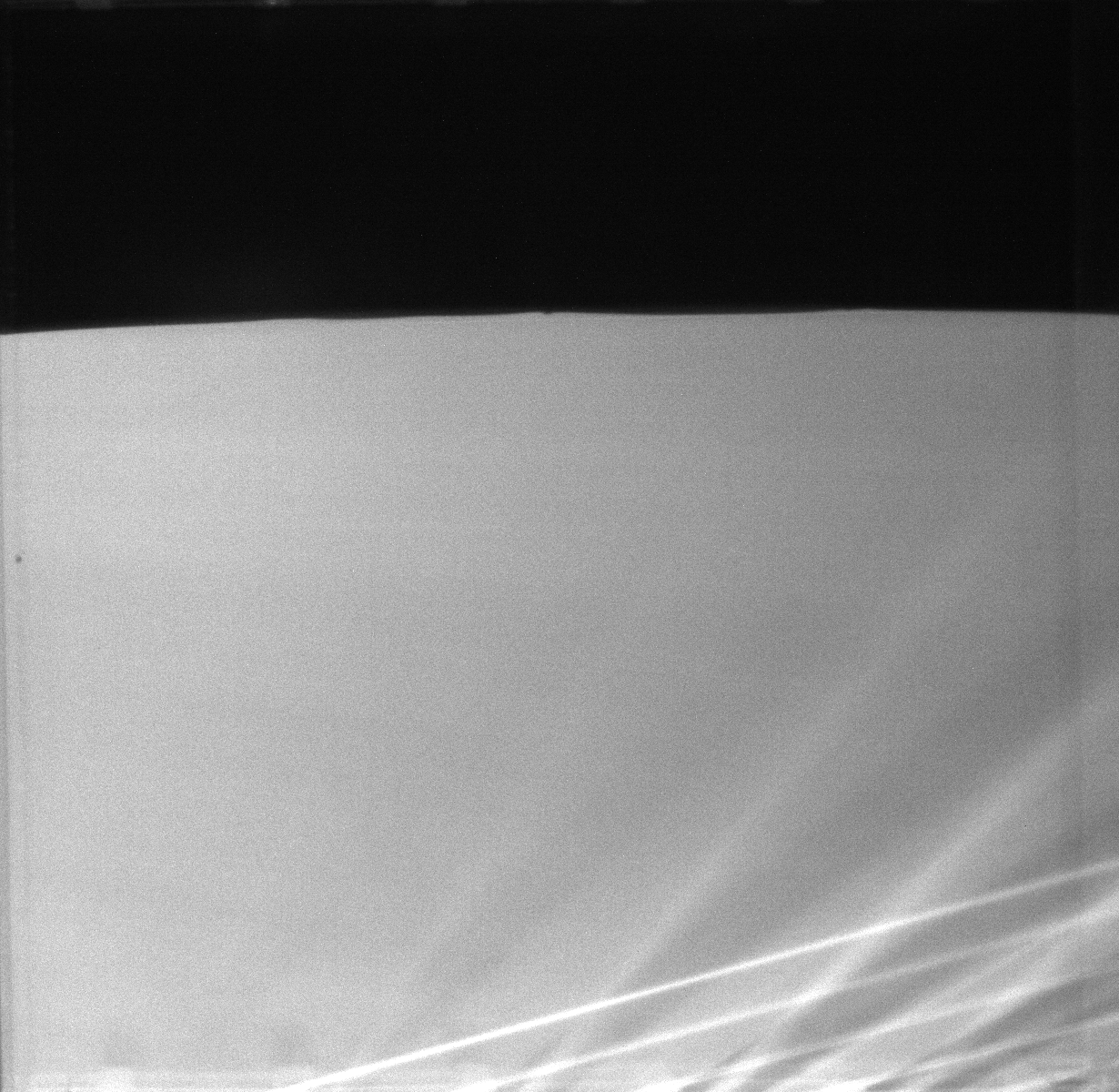}} \qquad
	\subfloat[t = 37 s.]{\includegraphics[width=0.25\textwidth]{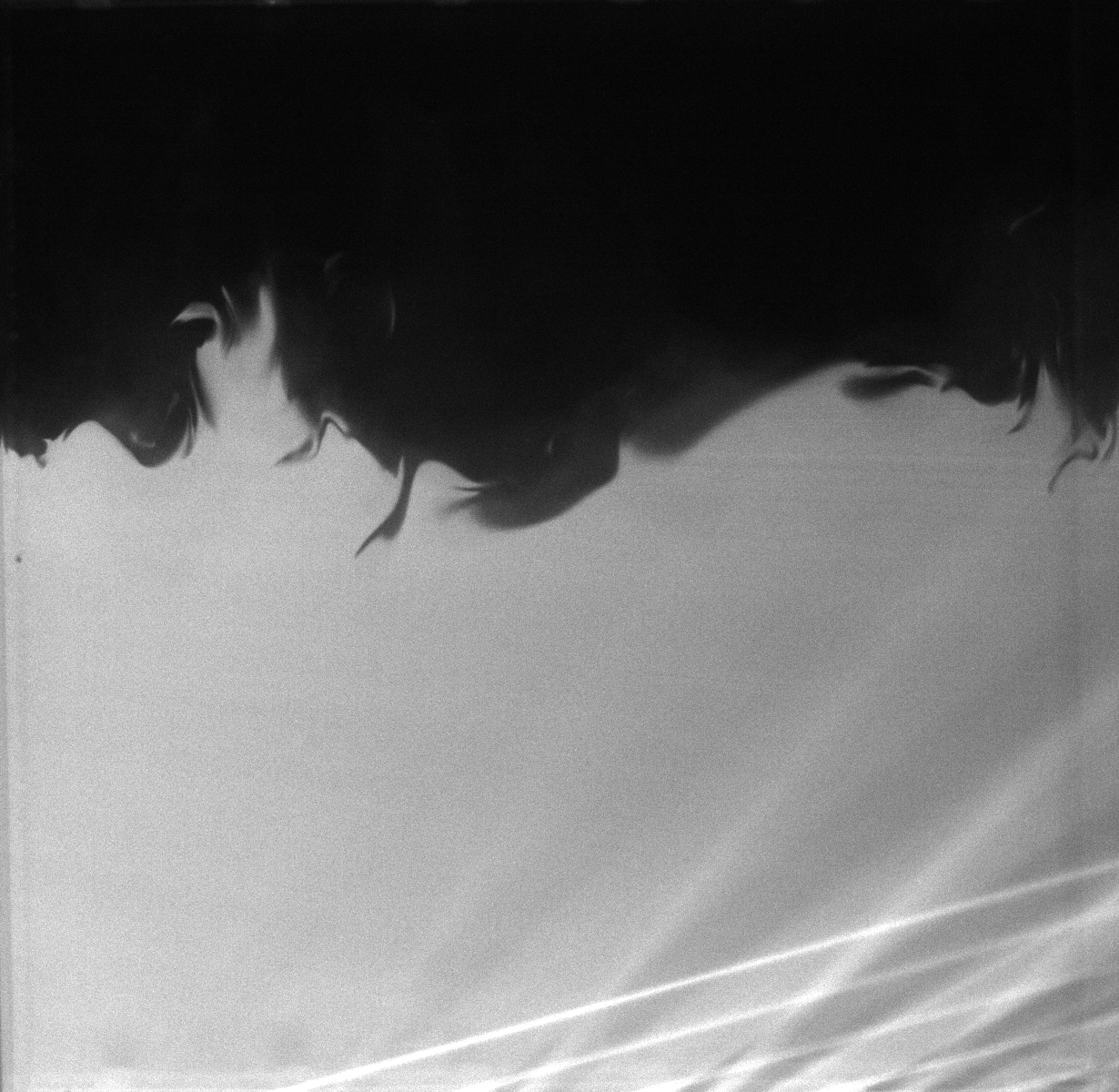}}
	\\
	\centering \textbf{Phase 1 -- Propagation}
	\vspace{0.2cm} \\
	\subfloat[t = 54 s.]{\includegraphics[width=0.25\textwidth]{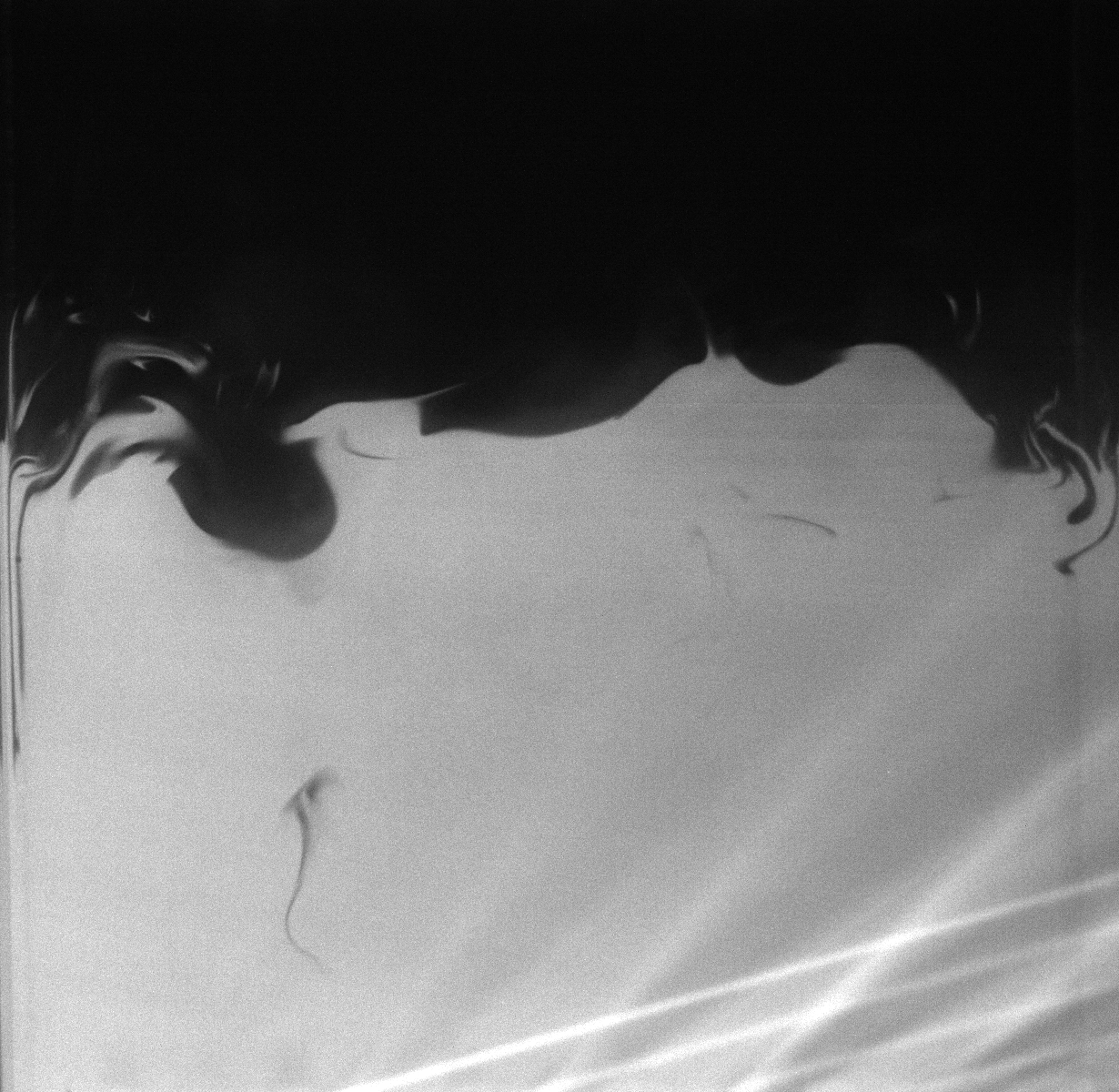}} \qquad
	\subfloat[t= 106 s.]{\includegraphics[width=0.25\textwidth]{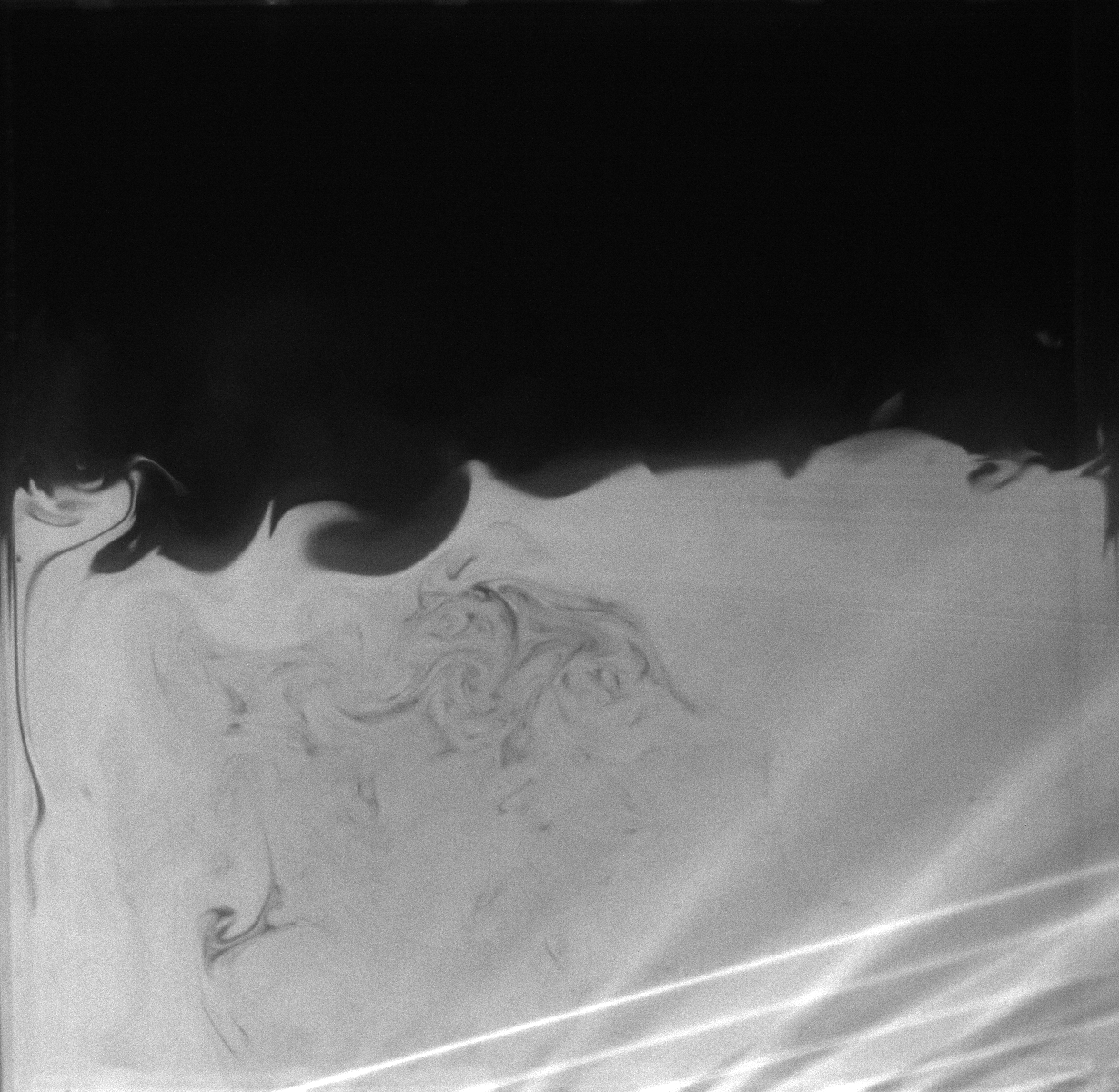}}
	\\
	\centering \textbf{Phase 2 -- Coexistence of propagation and mixing}
	\vspace{0.2cm} \\
	\subfloat[t = 187 s.]{\includegraphics[width=0.25\textwidth]{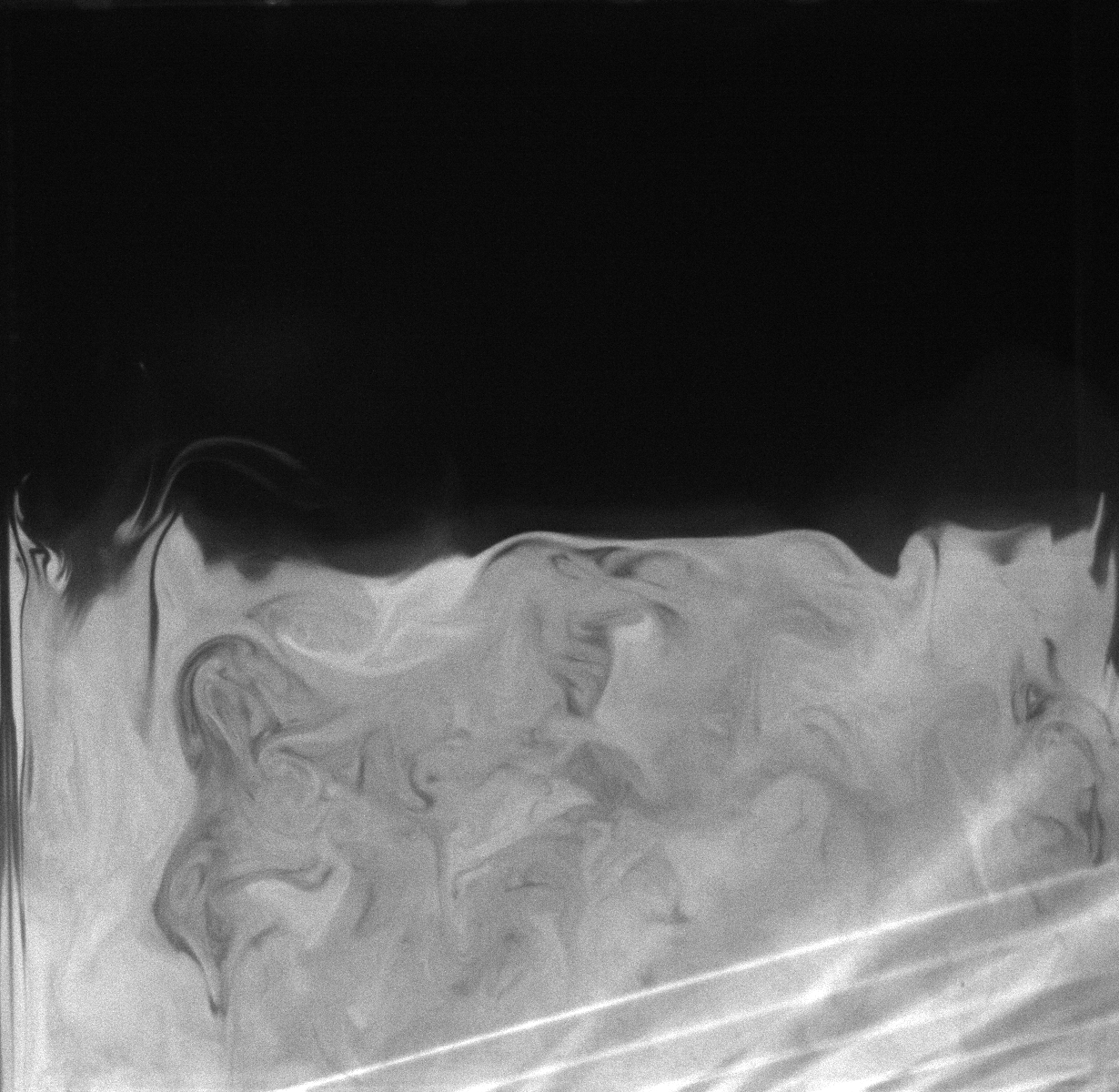}} \qquad
	\subfloat[t = 235 s.]{\includegraphics[width=0.25\textwidth]{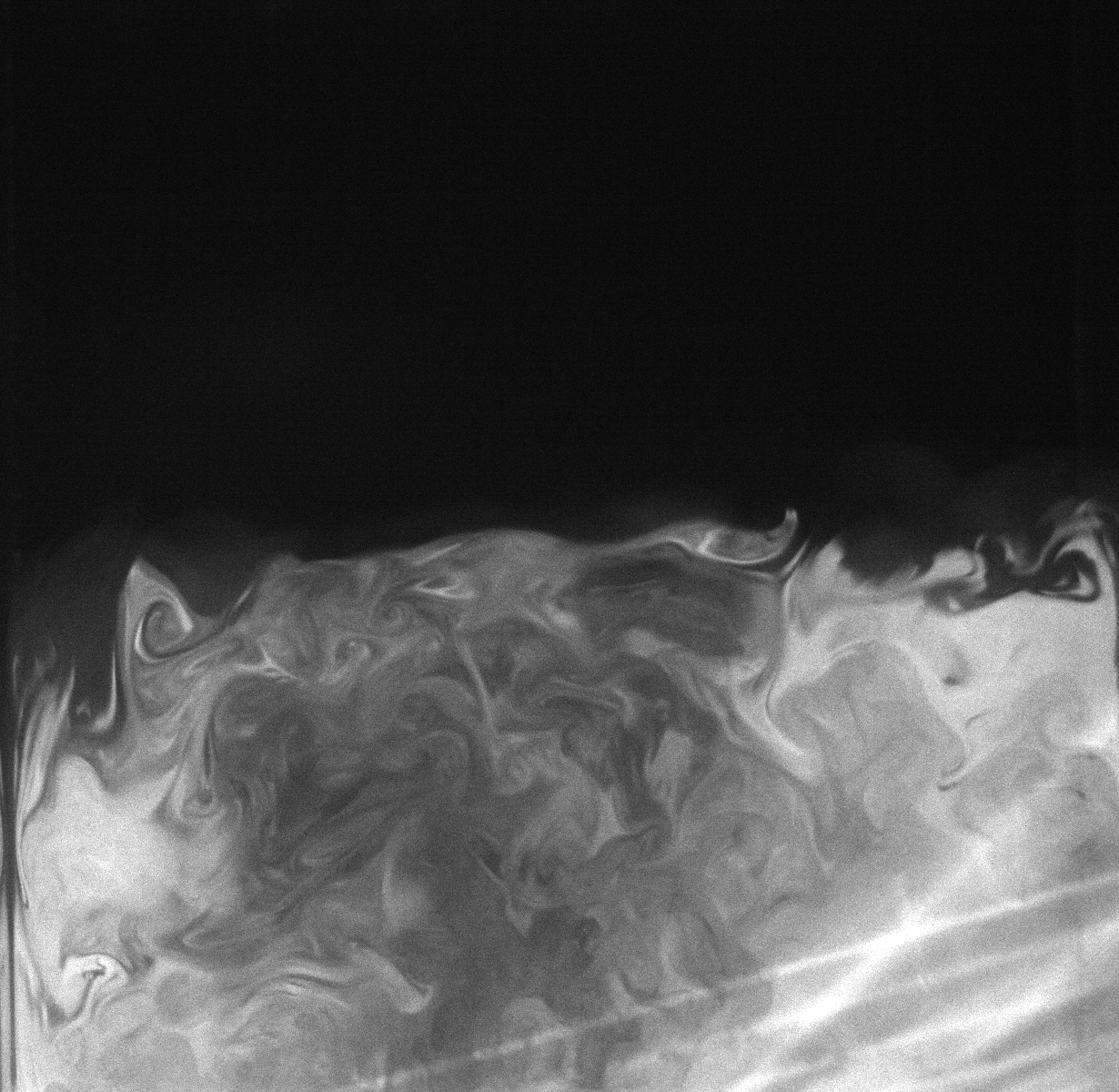}}
	\\
	\centering \textbf{Phase 3 -- Mixing phase}
	\caption{Snapshots of the front propagation with 2 oscillating grids highlighting the intermingle of the different regimes previously found, where $G=200$ mm, $A=10$ mm and $f=6$ Hz. The images are 20 cm wide and the lowest point of the upper grid and highest point of the lower grid are respectively located exactly at the top and bottom of the images.}
	\label{fig:G200}
\end{figure}

In the previous section, two different regimes were highlighted: a propagation regime when the oscillating grid is placed in the products, where the front is wrinkled and propagates from products to reactants; and 
a reactive mixing regime when the oscillating grid is placed in the reactants, where filaments of products are advected into the reactants and locally initiate new reactions in the bulk, accelerating the global reaction rate without  significant front displacement. When using both grids, one in the reactants and the second in the products, one thus expects a coexistence of both regimes. 
In order to qualitatively explore this case study, an experiment where the grids were as further apart a possible ($G = 200$~mm)  is shown in figure \ref{fig:G200}. After a laminar propagation phase, we start the grids oscillations at $t=0$ once the initially planar front is at $\sim 5$ cm from the upper grid. 
During the first minute, the front wrinkles and propagates towards the reactants, following a  propagating regime driven mainly by the upper grid. Then, black lamellae start appearing inside the reactants due to the bottom grid, while the front keeps advancing under the effect of the top grid. At this stage, both regimes coexist since both grids contribute to the front dynamics. When the front reaches just below the midpoint between the two grids, its propagation mostly stops and the reactive mixing phase dominates, leading to a strong acceleration of the global chemical reaction rate.

\begin{figure}[H]
	\centering
	\includegraphics[width=0.8\textwidth]{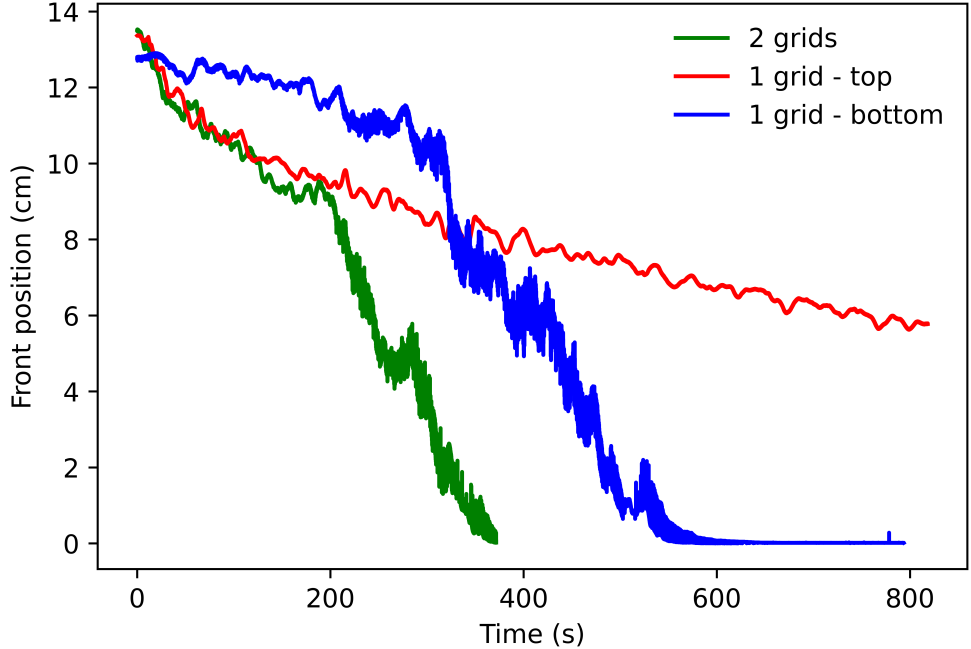}
	\vspace{-1cm}
	\caption{Time-evolution of the averaged front position oscillating the upper grid only (red), the bottom grid only (blue), and both grids with $G = 200$ mm (green). In the three experiments, $A = 10$ mm and $f = 6$ Hz.}
	\label{fig:2gVS1g}
\end{figure}

Figure \ref{fig:2gVS1g} compares the time evolution of the front position in experiments using both grids combined vs. each grid individually, confirming the observed succession of regimes over time. 
To obtain figure \ref{fig:2gVS1g}, we naively binarized the front images using a time-dependent threshold and then use equation (\ref{eq:AvFrontPos}). The curves resulting from the top-grid and 2-grids experiments collapse for the first $\approx 5$ cm, highlighting the influence of the upper grid. Then,  curves of the 2-grids and bottom-grid experiments bring out the same slope, thus showing the influence of the bottom grid. Note however that in this second stage, our measurement lacks physical meaning since the front is not a propagating interface anymore. Therefore in the following, to determine the turbulent propagation velocity $S_{T}$ when oscillating both grids, only the first part of the front position's time evolution - \textit{i.e.} in the propagation regime - is taken into account. 

\vspace{-1cm}

\subsection{Effect of turbulence and buoyancy on the front velocity}

\begin{figure}[H]
	\centering
	\includegraphics[width=\textwidth]{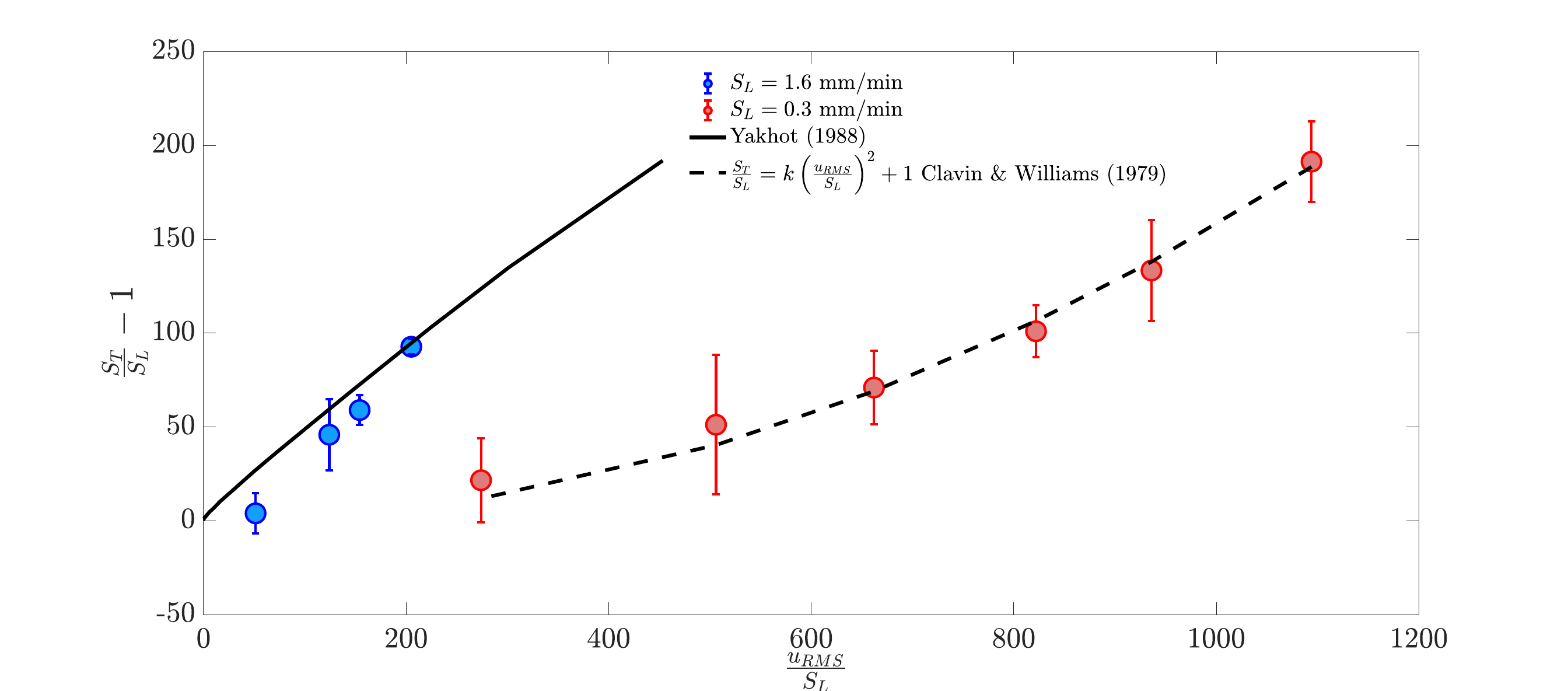}
	\caption{Turbulent front velocity as a function of the turbulence intensity in near homogeneous and isotropic turbulence, for {SOL1} (blue) and {SOL2} (red). Experiments at the lowest, highest and median turbulent intensities were repeated 3 times and the rest were repeated two times.}
	\label{fig:globalLaw2g}
\end{figure}

Figure \ref{fig:globalLaw2g} shows two series of experiments where $u_{\text{RMS}}$ in the ROI is varied by varying the grids control parameters, for two different laminar propagation velocities $S_{L}$. 
The turbulent front propagation velocity $S_{T}$, which in the selected cases is almost constant inside the ROI, is obtained by applying a linear fit over the time-evolution of the average front position computed using \eqref{eq:AvFrontPos}. Actually, for each result point, a series of linear fits were performed over progressively larger time intervals centered around the midpoint of the front temporal evolution: the error bars represent the standard deviation of all  calculated slopes. 

Figure \ref{fig:globalLaw2g} unravels different behaviors. 
First, for $S_{L} = 0.3$ mm/min (SOL2), the experimental data follow reasonably well a quadratic law just like in the single grid case, but for even higher values of $\frac{u_{RMS}}{S_{L}}$. The best fit of Clavin \& Williams' prediction \cite{clavin1979} gives a prefactor $k = 1.58 \cdot 10^{-4}$. 
On the other hand, experimental measurements for $S_{L}=1.6$~mm/min (SOL1) result in turbulent propagation velocities significantly larger at similar values of $u_{RMS}/S_L$. Rather than a quadratic law, these results seem to follow Yakhot's exponential law, in good agreement with \cite{shy1996a,shy1996b} who explored a range of  laminar front velocity including $S_{L}=1.6$~mm/min. 
Yakhot (1988) \cite{yakhot1988} derived his model from the equation of motion of a scalar field representing a flame surface, commonly called the $G-$equation, and found   
\begin{equation}
	\frac{S_{T}}{S_{L}} \sim \exp\left(\frac{u_{\text{RMS}}^{2}}{S_{T}^{2}}\right).
	\label{eq:yakhot}
\end{equation}
This law was validated in gas experiments using CH$_{4}$-air and C$_{2}$H$_{6}$-air for $1 < \frac{u_{\text{RMS}}}{S_{L}} \leq 20 - 25$, while $S_{T}$ was overestimated by \eqref{eq:yakhot} for higher turbulent intensities \cite{abdelgayed1984}. For a liquid flame, \cite{shy1992} noted that the validity of Yakhot's law requires $u_{\text{RMS}}/S_{L} \gtrsim 100$, which is inaccessible to gaseous flames. 

Actually, the strong discrepancy between our $S_{L}=0.3$ mm/min case and our $S_{L}=1.6$ mm/min case can only mean one thing: we are missing the influence of an additional  physical ingredient. From a dimensional point of view, it means that another typical velocity, in addition to $S_{L}$ (which characterizes the chemical kinetics) and $u_{\text{RMS}}$ (which characterizes the turbulence), plays a role in the overall propagation dynamic. A possible missing physical ingredient could be the viscosity and the associated viscous velocity $\nu/l$ giving rise, when compared to $u_{\text{RMS}}$, to the Reynolds number. But in \cite{shy1996a}, results have already shown independence on the Reynolds number. 
Besides, the viscosity of our 2 solutions is largely similar, so viscosity cannot explain their mismatch. Similarly, while the Schmidt number $Sc$ could have an influence on $S_{T}$, the typical diffusivity of the 2 considered solutions here is similar. So we argue that the missing key physical ingredient is the density difference between products and reactants, which varies from 0.05\% to 0.07\% in the two cases. 
Surely, it seems reasonable that the buoyancy jump has a strong influence on the front propagation: while turbulence wrinkles and stretches the front, buoyancy on the other hand tends to flatten it, therefore weakening the deformation. The Froude number $Fr$, defined as the ratio between $u_{\text{RMS}}$ and the typical phase velocity of gravity waves $\sqrt{gM\frac{\Delta \rho}{\rho}}$, quantifies the competition between both effects. In our experiments, $Fr$ remains of order of or smaller than 1, which indicates that despite a weak density difference, buoyancy forces are not negligible.

Therefore, rather than simply looking for a relation between $\frac{S_{T}}{S_{L}}$ and $\frac{u_{\text{RMS}}}{S_{L}}$, we shall include the possible influence of the Froude number $Fr$, hence look for instance for a scaling law of type 
\begin{equation}
	\frac{S_{T}}{S_{L}} = 1 + k\left(\frac{u_{\text{RMS}}}{S_{L}}\right)^{\alpha}Fr^{\beta}
	\label{eq:FrLaw}
\end{equation}
over the range of parameters studied here, i.e. for $Fr \lesssim \mathcal{O}(1) $ (we acknowledge that this scaling law cannot be generically valid since it diverges with $Fr$, i.e. in the absence of buoyancy effects).

\begin{figure}[H]
	\centering
	\includegraphics[width=0.8\textwidth]{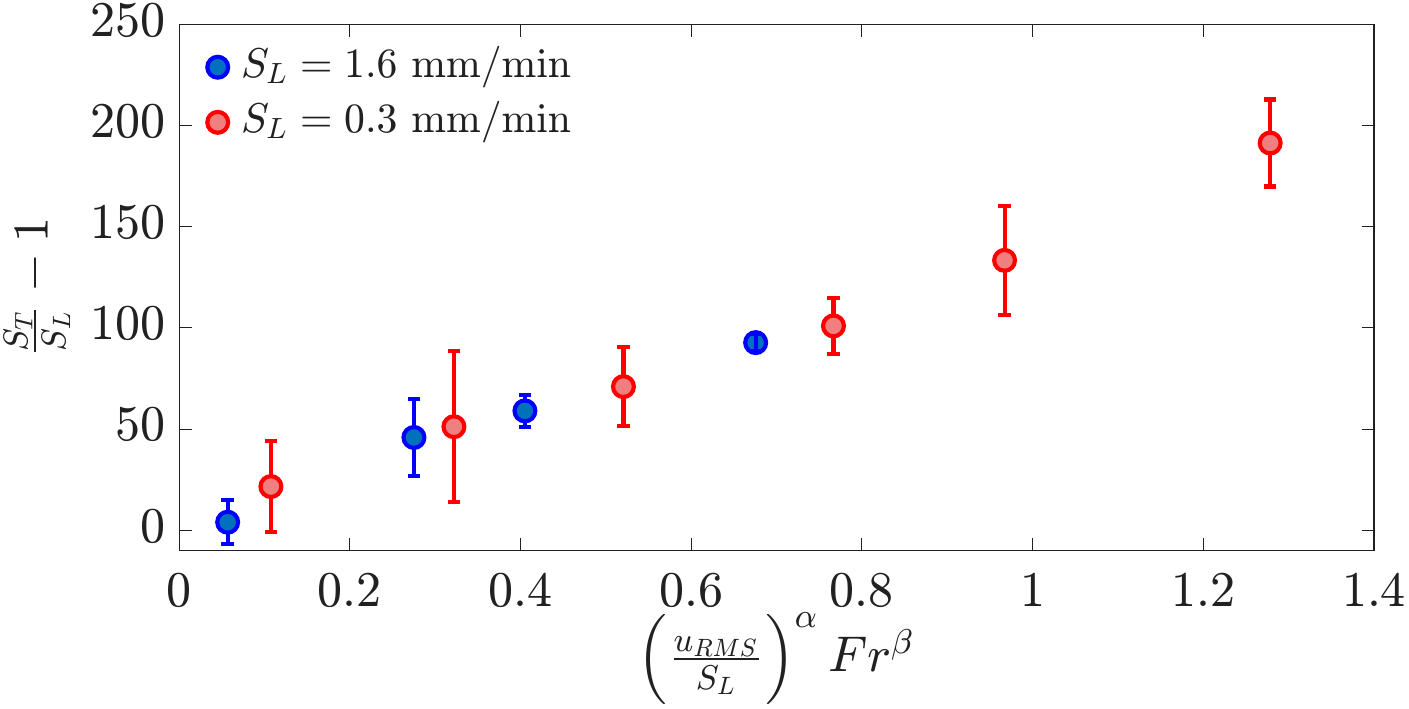}
	\caption{Non-dimensionalized front velocity $\frac{S_{T}}{S_{L}}$ as a function of  $\left(\frac{u_{\text{RMS}}}{S_{L}}\right)^{\alpha}Fr^{\beta}$ where $\alpha=0.22$ and $\beta=1.6$ were obtained by best fit.}
	\label{fig:FrLaw2g}
\end{figure}

In figure \ref{fig:FrLaw2g}, all the data points in figure \ref{fig:globalLaw2g} are plotted according to equation \eqref{eq:FrLaw}, over which the best fit on the power exponents gives $\alpha = 0.22$ and $\beta = 1.6$. Note that \eqref{eq:FrLaw} then retrieves an almost quadratic dependence on $u_{\text{RMS}}$ as previously discussed, since 
\begin{equation}
	\frac{S_{T}}{S_{L}} \propto \frac{u_{\text{RMS}}^{0.22}}{S_{L}^{0.22}}\frac{u_{\text{RMS}}^{1.6}}{\left(gM\frac{\Delta \rho}{\rho}\right)^{0.80}} \sim u_{\text{RMS}}^{1.82}.
\end{equation}
This scaling exponent is intermediate between the quadratic, deterministic regime predicted by Clavin and Williams \cite{clavin1979} and the stochastic $4/3$ regime of Kerstein and Ashurst \cite{kerstein1992}, both introduced in Section \ref{sec:Intro}. Indeed, although our flow is turbulent, gravity waves at the interface act as a regulatory mechanism, reducing the effective randomness of the velocity field experienced by the interface.

Now, regarding the scaling exponent in $Fr$, our value $\beta = 1.6$ can be compared with previous results for mixing across non-reactive density interfaces, e.g., two layers of fresh and salty water in a stable configuration, with turbulence generated by oscillating grids \cite{turner1968, hopfinger1975, poulainzarcos2022}, an incident plume \cite{baines1975}, or spherical vortices \cite{linden1974}. In those studies, the interface is progressively displaced away from the turbulent kinetic-energy source and the density contrast decreases as mixing occurs. An entrainment velocity $u_e$ is defined from the rate at which the interface deepens. Experimental and theoretical investigations generally report $u_e/u_{\text{RMS}} = K\,Fr^n$, where $K$ is non-universal and $n$ ranges between 1 and 3 depending on the setup and $Fr$ range (see e.g., discussion in\cite{herault2018erosion}).
In the present study, interface propagation is not driven solely by diffusive-eddy mixing and kinetic-to-potential energy conversion, but is also governed by the reaction. As a result, the two phases (products and reactants) remain separated, each with constant density, and the front velocity depends on both $u_{\text{RMS}}/S_L$ and $Fr$.

\begin{figure}[H]
	\centering
	\includegraphics[width=\textwidth]{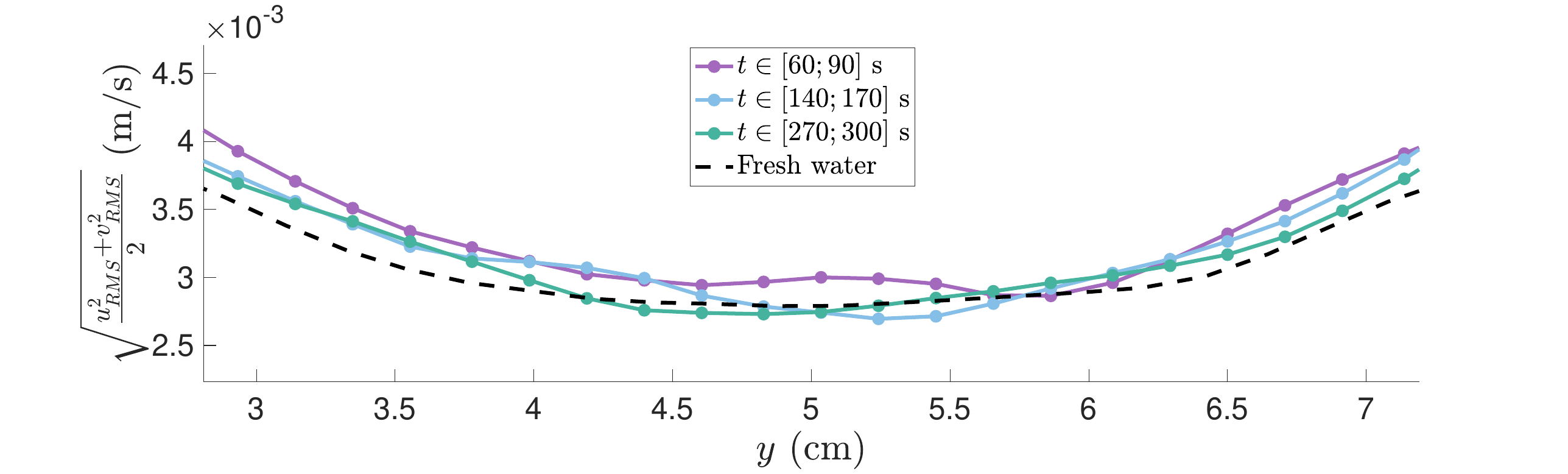}
	\caption{Velocity profiles along the vertical direction at different times during a reactive front propagation experiment inside a two-grid generated turbulence. The first grid is at $y= 0$ and the second at $y=10$ cm. Also shown for comparison is a mean profile in a fresh water experiment.} 
	\label{fig:velocity2g}
\end{figure}

Presumably the same role of buoyancy holds for the top-grid experiment, explaining the small jump between the two point series observed in figure \ref{fig:1glaw}.
However, the buoyancy effect is less pronounced in the single-grid case because our measure of the RMS velocity at the interface already largely takes it into account. This is not the case in the two-grid set-up where the density difference between reactants and products does not directly affect the velocity profile, since the turbulent energy is injected on both sides of the front. In figure \ref{fig:velocity2g},  velocity profiles in the ROI are plotted at 3 different times during a 6 minutes long experiment. Clearly, the front position is not distinguishable here, as opposed to the one-grid case shown in figure \ref{fig:u1g}. Furthermore, we retrieve the same profile with fresh water. So no clear feedback of the density jump between reactants and products is observed on the flow using two grids, and its whole influence on the front propagation velocity has to be accounted for independently by introducing $Fr$.

\section{Conclusion and prospects}

When looking at the growth of the turbulent front velocity as a function of turbulence intensity in gaseous combustion, various experimental, numerical and theoretical studies throughout the literature suggested a trend of the form $\frac{S_{T}}{S_{L}} \propto \left(\frac{u_{\text{RMS}}}{S_{L}}\right)^{\alpha}$, with $\alpha$ ranging from 0.5 to 2 \cite{clavin1979,michelson1977,williams1985,kerstein1988}. Yakhot's \cite{yakhot1988} exponential law was also found to be experimentally achieved at first-order, up until $\frac{u_{\text{RMS}}}{S_{L}} \approx 20$ \cite{yakhot1988}. For stronger turbulence, Huygens approximation breaks down as the reactive zone broadens and can no longer be described as a sharp interface. 

For a liquid flame, the Schmidt number is 500 times larger than in a gaseous flame, leading to momentum diffusing more rapidly compared to the species involved in the reaction hence resulting in a sharper and more well-defined reaction front even for even higher turbulent intensities. Moreover, slower laminar propagation velocities as well as laser-induced fluorescence allow significantly improved metrology, while suppressing compressibility effects allows to focus on the specific influence of turbulence. 
In this paper, we thus used a liquid flame for studying the evolution of the turbulent front velocity as a function of turbulence 
intensity. For turbulence generation, oscillating grids were used. Using a single grid generates a spatially decaying turbulence, for which two  regimes were unraveled. When the oscillating grid is located in the products, a propagation regime takes place, where the front remains well-defined and wrinkled under the effect of turbulence. In this configuration, the front propagation turned out to have a significant feedback on the flow arising from buoyancy effects induced by the density difference between products and reactants despite it being \textit{a priori} weak.  
When the grid is positioned on the reactants, catalyst-rich product filaments are advected inside the reactant, locally starting the reaction at different points of the bulk. In this mixing regime, the front is no longer localized, therefore surface creation is no longer the mechanism responsible for enhancing the reaction. Finally, using two out-of-phase oscillating grids gives rise to a central region where turbulence is nearly homogeneous and isotropic as found in the literature \cite{hopfinger1975, thompson1975, shy1997}. In our work, two different sets of chemical parameters were analysed. We found out that different laminar propagation velocities give rise to different tendencies of type $\frac{S_{T}}{S_{L}} = f\left(\frac{u_{\text{RMS}}}{S_{L}}\right)$, unraveling missing parameters to fully describe the behavior of turbulent flame propagation, other than $u_{\text{RMS}}$ and $S_{L}$. Following the noteworthy influence of buoyancy effect found in the previous setup, the missing parameter is suggested to be the Froude number $Fr$.

While these new conducted experiments have provided some insights on the role of turbulence and buoyancy in reactive fronts propagation, they have also raised a number of questions and opened up a vast array of possible follow-up studies. To delve deeper into the impact of $Fr$ and $\frac{u_{\text{RMS}}}{S_{L}}$, a larger range of reactants concentration - and therefore of $S_{L}$ and $\Delta \rho$ - is ought to be explored. 
And while this study primarily focused on the thin flame regime to remain in the framework of the Huygens approximation, higher turbulent intensities enable the study and quantification of the critical transition from thin to broadened reaction fronts in addition to determining the quenching limits, where turbulent agitation is so vigorous that the front ceases to propagate. By comparing the experimental results obtained from different reactant concentrations with the behavior of gaseous flames, these  findings can help to gain deeper insights into the fundamental processes governing turbulent flame propagation.

\nocite{townsend1957,hopfinger1973,turner1979,chu1976}

\end{document}